\documentclass[journal]{IEEEtran}

\ifCLASSINFOpdf

\else

\fi

\usepackage{float}
\usepackage{cite}
\usepackage{amsmath,amssymb,amsfonts}
\usepackage{algorithmic}
\usepackage{graphicx}
\usepackage{textcomp}
\usepackage{xcolor}
\usepackage{gensymb}
\usepackage{makecell}
\usepackage{stfloats}
\usepackage{amsthm}
\usepackage{amssymb}
\usepackage{amsmath}
\usepackage{mathtools}
\usepackage{extarrows}
\usepackage{bm}
\usepackage{mathrsfs}
\usepackage{stfloats}
\usepackage{cuted}
\usepackage{xpatch}
\usepackage{tabularx}

\usepackage{subfigure}

\begin{document}

\title{\textcolor{black}{FDA-MIMO-Based Integrated Multi-Target Sensing and Communication System with Complex Coefficients Information Embedding}}

\author{Jiangwei~Jian,~\IEEEmembership{Student Member,~IEEE,}
		Bang~Huang,
		Wenkai~Jia,
		Mingcheng~Fu,		
		Wen-Qin~Wang,~\IEEEmembership{Senior Member,~IEEE}
		and~Qimao~Huang
	\thanks{This work was supported by National Natural Science Foundation of China 62171092, and in part by the Postdoctoral Program for Innovation Talents under Grant BX20240054. (Corresponding author: Wen-Qin Wang).}
	\thanks{Jiangwei Jian, Wenkai Jia, Mingcheng Fu, and Wen-Qin Wang are with School of Information and Communication Engineering, University of Electronic Science and Technology of China, Chengdu, 611731, P. R. China. (Email: jianjiangwei@std.uestc.edu.cn; mrwenkaij@126.com; fumcheng@163.com; wqwang@uestc.edu.cn).
		
	Qimao Huang is with School of Physics, University of Electronic	Science and Technology of China, Chengdu, 611731, P. R. China. (Email: huangqimao@std.uestc.edu.cn). 
	
	Bang Huang is with the Computer, Electrical and Mathematical Science and Engineering (CEMSE) division in King Abdullah University of Science	and Technology (KAUST), Thuwal, Makkah Province, Saudi Arabia (Email: huangbang@std.uestc.edu.cn).
	}	
}


\maketitle

\begin{abstract}	
\textcolor{black}{The echo signals of frequency diverse array multiple-input multiple-output (FDA-MIMO) feature angle-range coupling, enabling simultaneous discrimination and estimation of multiple targets at different locations. In light of this, based on FDA-MIMO, this paper explores an sensing-centric integrated sensing and communication (ISAC) system for multi-target sensing. At the base station, we propose the FDA-MIMO-based spatial spectrum multi-target estimation (SSMTE) method, which first jointly estimates the angle and distance of targets and then estimates the velocities. To reduce the sensing computational complexity, the low-complexity spatial spectrum estimation (LCSSE) algorithm is proposed. LCSSE reduces the complexity without degrading the sensing performance by converting the joint angle-range search into two one-dimensional searches. To address the range ambiguity caused by frequency offset, a frequency offset design criterion (FODC) is proposed. It designs the integer and fractional components of the frequency offset to ensure the ambiguity distance exceeds the maximum sensing range, thereby alleviating parameters pairing errors. Moreover, the complex coefficients information embedding (CCIE) scheme is designed to improve system communication rates, which carries extra bits by selecting complex coefficients from the coefficient vector.} The closed-form expressions for the bit error rate (BER) tight upper bound and the Cramér-Rao bound (CRB) are derived. Simulation results show that the proposed system excels in multi-target sensing and communications.

\end{abstract}


\IEEEpeerreviewmaketitle

\section{Introduction}
\IEEEPARstart{T}{he} sharing of spectrum and hardware between radar and communications, termed integrated sensing and communication (ISAC), is emerging as a new trend in next-generation wireless networks \cite{gu2023integrated,ma2020joint,liu2022integrated}. On one hand, the ISAC technique enhances system spectrum efficiency and reduces hardware costs. On the other hand, it offers pervasive communication, sensing, and intelligence, fostering various emerging applications like autonomous driving, smart homes, and intelligent communications \cite{zheng2019radar,cui2021integrating,zhang2021overview}. Depending on differences in application focus, ISAC technology can be categorized into  sensing-centric, communication-centric, and communication-sensing trade-off designs \cite{liu2022integrated,di2013spatial,kumari2017ieee,liu2018mu}.

\textcolor{black}{The sensing-centric design regards the targets sensing performance as the primary function, which is mainly used in scenarios such as auto-driving, UAV surveillance and geo-mapping \cite{liu2022integrated,ma2020joint}.} As early as 1960, \cite{mealey1963method} proposed an ISAC system by embedding communication data within radar pulse intervals. However, the phased-array radar used lacked waveform diversity gain. To address this, multiple-input multiple-output (MIMO)-based ISAC systems with waveform diversity degrees of freedom (DoFs) have garnered widespread attention \cite{hassanien2015dual,10770015,yu2022integrated,wang2018dual,huang2020majorcom,ma2021frac}. Specifically, \cite{hassanien2015dual} designed the mainlobe of the MIMO radar to sense target, and the sidelobe to communication. \textcolor{black}{To improve the spectrum utilization while maintaining sensing performance, index modulation (IM)-aided ISAC techniques attracted researchers' attention \cite{10770015}. In this regard, \cite{yu2022integrated} embedded communication data into spatio-spectral passbands and stopbands, and optimized the beampattern to guarantee the sensing performance. Besides,\cite{wang2018dual} proposed activating partial transmit antennas to carry additional index bits, achieving target perception through sparse array. As a further study, an ISAC system based on frequency agile radar was proposed in \cite{huang2020majorcom}, which utilized frequency offset selection to convey index bits and proposed the multi-target sensing method. Later, \cite{ma2021frac} extended this work by jointly IM in frequency and spatial dimensions, which further improved the multi-target estimation accuracy via compressed sensing.}

On the flip side, the communication-centric and trade-off designs focus on the high communication performance and sensing-communication balance, respectively. In terms of communication-centered design, in \cite{mizui1993vehicle}, the sensing function was attached to the spread spectrum communication systems to realize the dual function. Motivated by this, the orthogonal frequency division multiplexing (OFDM) based ISAC system was proposed in \cite{sturm2011waveform}, which estimates targets by processing the OFDM echo signals in the fast-slow time domain. Later, MIMO was combined with OFDM, proposed as MIMO-OFDM-based ISAC systems, to improve the communication and sensing performance. These encompass subcarriers allocation and optimization \cite{xu2023bandwidth}, channel interference exploitation \cite{keskin2021mimo}, precoding \cite{temiz2021optimized}, waveform optimization \cite{johnston2022mimo}, and uplink design \cite{temiz2021dual}. In the trade-off design, the focus lies in optimizing joint sensing and communication metrics through precoder design. This includes optimizing metrics such as the users sum rate and sensing beampattern optimization \cite{liu2018toward,liu2022transmit}, Cramér-Rao bound (CRB) and communication signal-to-interference-plus-noise ratio (SINR) optimization \cite{liu2021cramer}.

However, aforementioned ISAC systems mainly relied on phased arrays (PAs), whose   steering is only angle-dependent without range information. An emerging frequency diverse array (FDA)-MIMO technique extends the DoFs of signal processing to the angle-range dimension by introducing a frequency offset among the adjacent elements \cite{sammartino2013frequency,huang2022adaptive,jia2023optimal}. Inspired by this, FDA-MIMO radar has been applied in high-resolution target estimation \cite{lan2021single,xu2015joint}, target detection \cite{lan2020glrt,gui2020low}, range clutter suppression \cite{sun2024space}, mainlobe interference suppression \cite{lan2020suppression,xu2015deceptive}, and exhibited superior radar performance to the PA-based MIMO in the range dimension. \textcolor{black}{Moreover, FDA-MIMO can also benefit communications. \cite{cheng2021physical,jian2023physical} described how the angle-range coupling character of FDA was utilized to guarantee communication security for specific location users. \cite{wu2021high} demonstrated the FDA's beam-tracking advantage in highly dynamic communications. \cite{jian2023mimo} employed the frequency offset of FDA as IM entities, further enhancing the communication rates and bit error rate (BER) performance.}

FDA-MIMO has demonstrated attractive performance in both radar and communications, driving its incorporation into ISAC systems \cite{nusenu2018time,wu2023waveform,zhou2021performance}. Specifically, \cite{nusenu2018time} embedded communication bits into the spreading sequence of each pulse, yielded satisfactory sensing performance. Another approach involved embedding constellation symbols into multiple sub-pulses witin one pulse, enabling simultaneous communication and sensing \cite{wu2023waveform}. Further, \cite{zhou2021performance} proposed embedding phase modulation symbols into FDA-MIMO radar waveforms and optimized the transmit beamforming to achieve the sensing-communication performance balance. Moreover, \cite{li2023joint} convey extra bits by permutating the transmit frequency offsets, which improved communication rates and CRB performance. However, the challenges of enhancing communication rates and accurately estimating targets persist. To address this, \cite{jian2023fda} proposed the frequency offset permutation index modulation (FOPIM) scheme, which involved selecting and permutating frequency offsets to carry additional bits, along with the target estimation method. This approach achieved superior sensing performance compared to MIMO-based ISAC systems.

Nevertheless, the aforementioned FDA-MIMO-based ISAC system failed to consider the multi-target sensing, and how to suppress the range periodicity during multi-target estimation remains an open question. Moreover, the FOPIM method only activates partial frequency offsets, leading to the spectrum wastage. Motivated by this, this paper explores the FDA-MIMO-based ISAC system in multi-target scenarios and proposes a complex coefficients information embedding (CCIE) transmission scheme independent of frequency offsets. The main contributions of our work are listed as follows:
\begin{itemize}
\item [1)]
\textcolor{black}{In this work, a spatial spectrum multi-target estimation (SSMTE) method is proposed for multi-target sensing. Specifically, within the target-containing range bins, the angles and ranges of targets are jointly estimated in the spatial spectrum of FDA-MIMO. Subsequently, the least squares (LS) is employed to estimate the velocity.} The SSMTE method suffers from high complexity due to its angle-range two-dimensional (2-D) search. To tackle this issue, the low-complexity spatial spectrum estimation (LCSSE) approach is proposed, which dramatically reduces the complexity by converting the 2-D angle-range search into two one-dimensional (1-D) searches. Simulation results show that LCSSE and SSMTE methods have similar sensing performance.

\item [2)]
The FDA-MIMO exhibits periodic variation in its steering vector with range, resulting in range ambiguity in target estimation. To tackle this issue, the frequency offset design criterion (FODC) is designed in this paper. FODC designs the integer and fractional components of each transmit frequency offset to ensure that the range periodicity of the steering vector exceeds the maximum sensing distance, thereby mitigating range ambiguity in multi-target estimation. Moreover, we derive closed-form expressions for the system CRB performance.

\item [3)]
\textcolor{black}{We propose the CCIE scheme to enhance the communication rate. In the CCIE method, each antenna selects a complex coefficient from a normalized complex coefficient vector to transmit additional bits and conveys an independent quadrature amplitude modulation (QAM) symbol.} Additionally, the closed-form expressions for the system BER tight upper bound are derived to evaluate the communication performance.

\end{itemize}	

The rest of this paper is organized as follows. Section \ref{sc2} proposes the CCIE approach for the FDA-MIMO-based ISAC system. Section \ref{sc3} discusses the signal processing of system sensing and communication receivers. Section \ref{SPA} analyzes the theoretical performance of the system CRB, complexity and BER. Finally, simulation results are discussed in section \ref{sc5}, and the paper is concluded in section \ref{sc6}.

$Notations$: $^T$, ${^{\ddagger}}$ and $^{\dagger}$ stand for the transpose, conjugate and conjugate transpose operations, respectively. $\mathbf{I}_N$ denotes the identity matrix of order $N$. $\lfloor \cdot \rfloor$, $!$ and $\Gamma (\cdot)$ denote the floor function, factorial and Gamma function, respectively. $\mathrm{Tr}[\cdot] $ represents the trace operation and $j=\sqrt{-1}$. $\mathrm{Re}\left\{ \cdot \right\}$ and $\mathrm{Im}\left\{ \cdot \right\}$ are the real part and the imaginary part operators, respectively. $\odot$ and $\otimes$ stand for the Hadamard product and Kronecker product operations, respectively. $\lfloor \cdot \rfloor _{LCM}$ denots the least common multiple operation. $*$ represents the convolution operation. $\mathrm{diag}()$ denotes the vector diagonalization operation.

\section{System Model}
\label{sc2}
\begin{figure}[htp]
	\vspace{-0.3cm} 
	\centering
	\subfigure	{\includegraphics[width=0.35\textwidth]{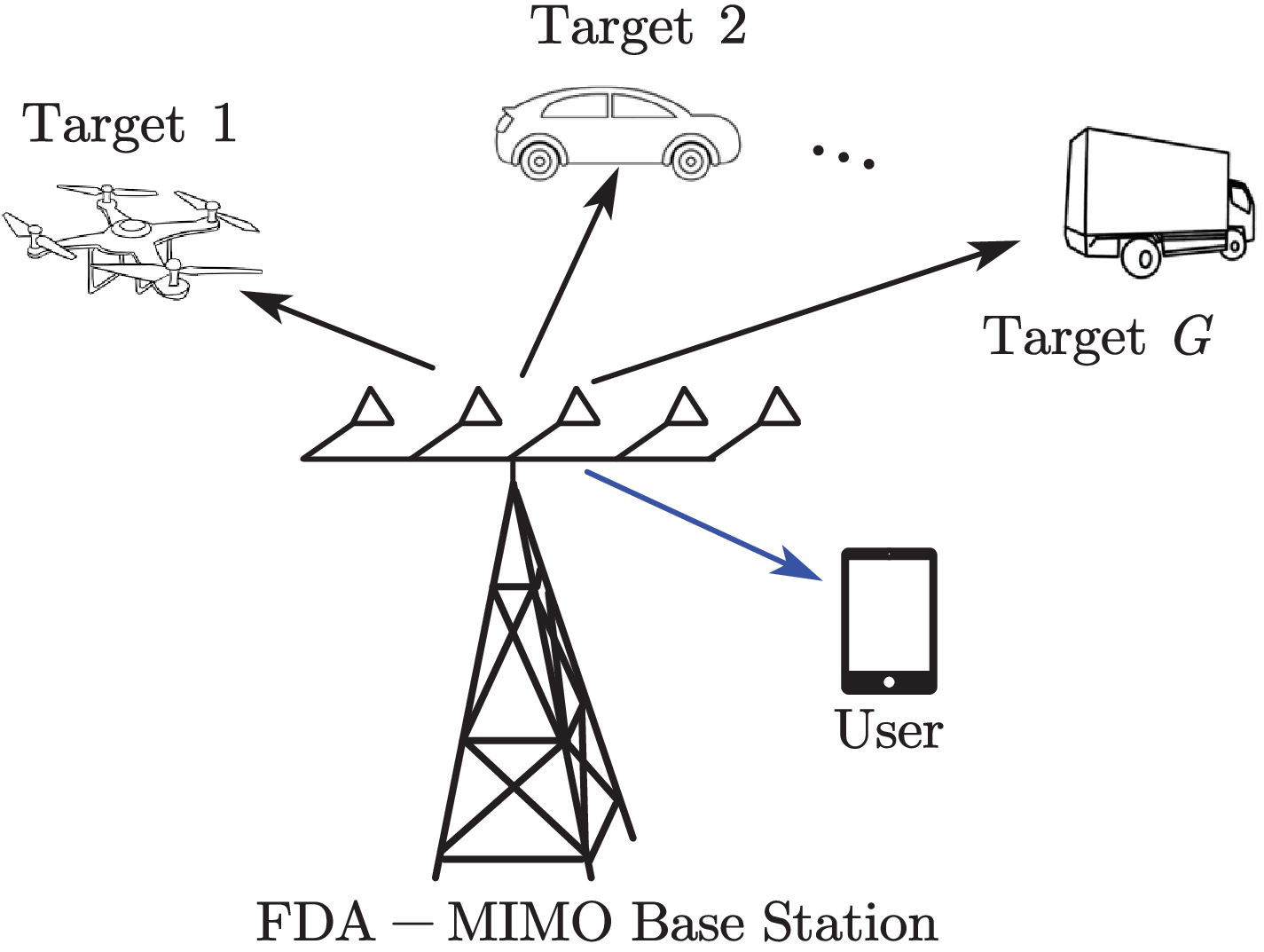}}
	\caption{System model.}
	\label{system_model}
\end{figure}

This paper considers an ISAC system as shown in Fig. \ref{system_model}. The FDA-MIMO base station (BS) equips $N$ transmit antennas and $M$ receive antennas for sensing $G$ targets, while serving a communication user equipped $U$ antennas. On the transmitter side of ISAC systems, IM techniques are widely adopted to enhance the system communication rates \cite{ma2021frac,huang2020majorcom,yu2022integrated}. Although some recent works have combined FDA with IM for ISAC systems, they carried additional information by activating some frequency offsets \cite{jian2023mimo,jian2023fda,ma2021frac}, which resulted in a waste of spectrum resources. To overcome this drawback as well as to further enhance the communication rate, the CCIE method is proposed in this paper.

The proposed CCIE scheme carries extra bits by combining the constellation symbols with elements in a complex coefficient vector. Specifically, we generate a normalized complex coefficient vector $\boldsymbol{c}=[ c_1,\cdots ,c_j,\cdots ,c_J ] ^T\in \mathbb{C} ^{J\times 1}$ with $\frac{\boldsymbol{c}^{\dagger}\boldsymbol{c}}{J}=1$, where the elements are shared with the communication user. We define a $K$ pulse repetition interval (PRI) as a coherent processing interval (CPI) and each PRI is of length $T$. In the $k$th PRI, the trasmitted constellation symbol vector is denoted as $\mathbf{x}_k=[ x_{1}^{k},\cdots ,x_{n}^{k},\cdots ,x_{N}^{k} ] ^T\in \mathbb{C} ^{N\times 1}$, where $x_{n}^{k}$ denotes the $L$-ary unit energy QAM symbol. Then, the CCIE symbol is designed as
\begin{equation}\label{x_n_k}
	\begin{aligned}
		\tilde{x}_{n}^{k}=\boldsymbol{c}^{T}\boldsymbol{\varsigma }_{n}^{k}=c_{i_{n,k}}x_{n}^{k},
	\end{aligned}
\end{equation}
where 
\begin{equation}\label{var_n_k}
	\begin{aligned}
		\boldsymbol{\varsigma }_{n}^{k}=[ 0,\cdots ,\underset{i_{n,k}\mathrm{th}}{\underbrace{x_{n}^{k}}},\cdots ,0 ] ^T\in \mathbb{C} ^{J\times 1},
	\end{aligned}
\end{equation}
denotes the complex coefficient selection vector of the $n$th antenna at the $k$th PRI. $c_{i_{n,k}}$ stands for the complex coefficient selected from $\boldsymbol{c}$, whereas $i_{n,k}$ representing the index of $c_{i_{n,k}}$ in $\boldsymbol{c}$. From \eqref{x_n_k} and \eqref{var_n_k}, we can claim that the proposed CCIE method can carry $N\times (\lfloor \log _2J \rfloor +\log _2L)$ bits in one transmission.

The transmitter, namely FDA-MIMO BS, is considered as a uniform linear array (ULA). The transmit frequency of the $n$th BS antenna is designed as
\begin{equation}\label{f_ns}
	\begin{aligned}
		f_n=f_c+(n-1) \Delta f,
	\end{aligned}
\end{equation}
where $f_c$ denotes the common carrier frequency, whereas $\Delta f$ denotes the frequecy offset increment. Following the proposed CCIE scheme, the transmit signal of the $n$th antenna at the $k$th PRI is expressed as 
\begin{equation}\label{psis}
	\begin{aligned}
		s_{n}^{k}(t) =\varrho (t-kT) \tilde{x}_{n}^{k}e^{j2\pi [ f_c+(n-1) \Delta f ] t}, 
	\end{aligned}
\end{equation}
where $k=0,1,\cdots ,K-1$. $\varrho (t)$ is the unit energy baseband waveform with the pulse duration $T_W$, which satisfies the following orthogonality \cite{gui2020low,lan2020suppression}:
\begin{equation}\label{orthogonality}
	\begin{aligned}
		\int_{-\infty}^{\infty}{\varrho (t)\varrho ^{\dagger}(t-\tau )e^{j2\pi \left( m-m\prime \right) \Delta ft}}dt=\left\{ \begin{array}{c}
			1,m=m'\\
			0,m\ne m'\\
		\end{array} \right. ,\forall \tau.
	\end{aligned}
\end{equation}

\section{System communication and sensing functions}
\label{sc3}
In this section, we model the received signals of the communication and sensing receivers, as well as design signal processing methods.

\subsection{Sensing receiver}
We assume that the locations of $G$ point targets in Fig. \ref{system_model} is $\left\{ (R_1,\theta _1) ,\cdots ,(R_g,\theta _g) ,\cdots ,(R_G,\theta _G) \right\} $ and the propagation path with the base station is the line-of-sight \cite{sturm2011waveform,ma2021frac,liu2022transmit,xu2023bandwidth}. Then, on the BS side, the received signal of the $m$th antenna in the $k$th PRI can be written as
\begin{equation}\label{S_nk}
	\begin{aligned}
		y_{m}^{k}\left( t \right) =&\sum_{n=1}^N{\sum_{g=1}^G{\xi _ge^{j2\pi \mathcal{F} _gt}s_{n}^{k}(t-\tau _{m,n,g})}}
		\\
		\approx& \sum_{n=1}^N{\sum_{g=1}^G{\xi _ge^{j2\pi \mathcal{F} _gt}\varrho  (t-kT-\tau _g)\tilde{x}_{n}^{k}e^{j2\pi (f_c+\Delta f_{n}) t}}}
		\\
		&\times e^{-j2\pi \Delta f_{n}\frac{2R_g}{c}}e^{j2\pi \frac{f_c(n-1) d_1\sin \theta _g}{c}}e^{j2\pi \frac{f_c( m-1) d_2\sin \theta _g}{c}},
	\end{aligned}
\end{equation}
where $\Delta f_{n}=(n-1)\Delta f$ denots the frequency offset of the $n$th transmit antenna. $\tau _{m,n,g}=\frac{2R_g-(n-1) d_1\sin \theta _g-(m-1) d_2\sin \theta _g}{c}$ represents the delay between the $n$th transmit antenna and the $m$th receive antenna for the $g$th target. $d_1$ and $d_2$ denote the spacing of neighboring elements in the transmit and receive arrays, respectively. $c$ represents the light speed. $\xi _g$ is the reflection coefficient of the $g$th target, which absorbs the constant term $e^{-j2\pi f_c\frac{2R_g}{c}}$ \cite{xu2015joint}. Note that the approximation $\varrho (t-kT-\tau _{m,n,g}) \approx \varrho (t-kT-\tau _g)$ is considered in \eqref{S_nk} under the narrow-band assumption. The terms $e^{j2\pi \Delta f_{n}\frac{(n-1) d_1\sin \theta _g}{c}}$, $e^{j2\pi \Delta f_{n} \frac{(m-1)d_2\sin \theta _g}{c}}$ are tiny enough to be ignored \cite{lan2020glrt}. $\mathcal{F} _g\approx \frac{2v_g}{c}f_c$ and $v_g$ denote the doppler shift and velocity of the $g$th target, respectively. Note that, similar to \cite{xu2023bandwidth,gui2018general}, the doppler spreading from the frequency offset is ignored in this paper.

\begin{figure}[htp]
	\vspace{-0.3cm} 
	\centering
	\subfigure	{\includegraphics[width=0.45\textwidth]{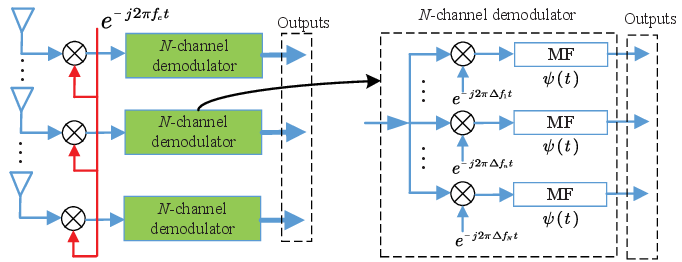}}
	\caption{Structure of the sensing and communication receivers.}
	\label{receiver_struc}
	\vspace{-0.2cm} 
\end{figure}
The sensing receiver structure is shown in Fig. \ref{receiver_struc}, which can also be deployed to the communication user without additional design. The received signal is first down-converted with $e^{-j2\pi f_ct}$, and followed by the $N$-channel demodulator. In the $n$th channel, the down-converted signal is multipled by $e^{-j2\pi \Delta f_nt}$ and then match-filtered by $\varrho (t)$. Following this, the filtered signal of the $n$th channel of the $m$th receive antenna can be expressed as \cite{ma2021frac}
\begin{equation}\label{y_mk}
	\begin{aligned}
		y_{m}^{k}=& y_{m}^{k}(t) e^{-j2\pi f_ct}e^{-j2\pi \Delta f_nt} *\varrho (t)
		\\
		=&\sum_{g=1}^G{\xi _ge^{j2\pi \psi _{g}^{k}}\tilde{x}_{n}^{k}}e^{-j2\pi (n-1) \Delta f\frac{2R_g}{c}}
		\\
		&\times e^{j2\pi \frac{f_c(n-1) d_1\sin \theta _g}{c}}e^{j2\pi \frac{f_c(m-1) d_2\sin \theta _g}{c}}.
	\end{aligned}
\end{equation}
where $\psi _{g}^{k}=\mathcal{F} _g (k-1) T$ \cite{gui2018general}. Then, the $N$ outputs of the $m$th receive antenna can be stacked into a vector as
\begin{equation}
	\begin{aligned}
		\mathbf{y}_{m}^{k}&=[ y_{m,1}^{k},\cdots ,y_{m,n}^{k},\cdots ,y_{m,N}^{k} ] ^T
		\\
		&=\sum_{g=1}^G{\xi _ge^{j2\pi \psi _{g}^{k}}\tilde{\mathbf{x}}_{}^{k}\odot \mathbf{a}_T (R_g) \odot \mathbf{a}_T(\theta _g) e^{j2\pi \frac{f_c (m-1) d_2\sin \theta _g}{c}}},
	\end{aligned}
\end{equation}
where 
\begin{equation}\label{a_T_R_g}
	\begin{aligned}
		\mathbf{a}_T(R_g) =\left[ 1,\cdots,e^{j\varphi _n(R_g)},\cdots ,e^{j\varphi _N(R_g)} \right] ^T,
	\end{aligned}
\end{equation}
and
\begin{equation}\label{a_T_theta}
	\begin{aligned}
		\mathbf{a}_T(\theta _g) =\left[ 1,\cdots,e^{j\phi _n(\theta _g)},\cdots ,e^{j\phi _N(\theta _g)} \right] ^T,
	\end{aligned}
\end{equation}
represent the transmit range and angle steering vectors, respectively. Note that $\varphi _n(R_g) =-2\pi \Delta f_n\frac{2R_g}{c}$ and $\phi _n( \theta _g) =2\pi \frac{f_c(n-1) d_1\sin \theta _g}{c}$ in \eqref{a_T_R_g} and \eqref{a_T_theta}. $\tilde{\mathbf{x}}_{}^{k}=[ \tilde{x}_{1}^{k},\tilde{x}_{2}^{k},\cdots ,\tilde{x}_{N}^{k}] ^T$ denotes the emitted CCIE symbol vector.

Further, the demodulated outputs of all channels of the $M$ antennas can be written as
\begin{equation}\label{yk_vec}
	\begin{aligned}
		\mathbf{y}^{k}&=\left[ {\mathbf{y}_{1}^{k}}^T,\cdots ,{\mathbf{y}_{m}^{k}}^T,\cdots ,{\mathbf{y}_{M}^{k}}^T \right] ^T
		\\
		&=\sum_{g=1}^G{\xi _ge^{j2\pi \psi _{g}^{k}}\mathbf{a}_R\left( \theta _g \right) \otimes \left[ \tilde{\mathbf{x}}^{k}\odot \mathbf{a}_T\left( R_g \right) \odot \mathbf{a}_T\left( \theta _g \right) \right]}+\mathbf{n}_k,
	\end{aligned}
\end{equation}
where 
\begin{equation}\label{a_R_theta}
	\begin{aligned}
		\mathbf{a}_R(\theta _g) =\left[ 1,\cdots ,e^{j\omega _m\left( \theta _g \right)},\cdots ,e^{j\omega _M(\theta _g)} \right] ^T,
	\end{aligned}
\end{equation}
stands for the receive steering vector with $\omega _m(\theta _g) =2\pi \frac{f_c(m-1) d_2\sin \theta _g}{c}$. $\mathbf{n}_k\sim \mathcal{C} \mathcal{N}  (0,\sigma _{1}^{2}\mathbf{I}_{M\times N}) $ represents the receive noise vector with $\sigma _{1}^{2}$ denoting the noise power.

From \eqref{yk_vec}, one can observe that the communication symbol term, $\tilde{\mathbf{x}}^{k}$, degrades the sensing performance. To address this dilemma, we introduce a sensing compensation vector, as
\begin{equation}\label{b_R_comp}
	\begin{aligned}
		\mathbf{b}_R=\mathbf{1}_M\otimes \tilde{\mathbf{x}}_{}^{k},
	\end{aligned}
\end{equation}
where $\mathbf{1}_M\in \mathbb{C} ^{M\times 1}$ denotes the all-one vector. Consequently, the compensated receive signal in the $k$th PRI can be expressed as 
\begin{equation}\label{y_tilde}
	\begin{aligned}
		\tilde{\mathbf{y}}_{}^{k}&=\mathbf{y}_{}^{k}\odot \mathbf{b}_R
		\\
		&=\sum_{g=1}^G{\xi _ge^{j2\pi \psi _{g}^{k}}\mathbf{a}_R( \theta _g ) \otimes [ \mathbf{a}_T( R_g ) \odot \mathbf{a}_T( \theta _g ) ]}+\mathbf{n}_k.
	\end{aligned}
\end{equation}

Equations \eqref{b_R_comp} and \eqref{y_tilde} show that at the sensing receiver, the interference of communication symbols can be removed by compensating the received data using prior communication information. In other words, the proposed system does not need to consider communication and sensing balance. This is facilitated by the orthogonality between the transmitted waveforms, which allows the receiver to process the data from each demodulation channel separately. Similar methods are reported in \cite{sturm2011waveform,ma2021frac}.

In the sequel, our focus turns to the estimation of range, angle, and velocity. Although conventional MIMO radars are capable of angle, range, and velocity estimations, the issue of how to pair the estimated parameters remains open. The FDA-MIMO can fill this gap well. From \eqref{y_tilde}, one can observe that the range and angle parameters are coupled \cite{xu2015joint,lan2020suppression}. With this observation, this paper proposes the FDA-MIMO-based SSMTE method. 

Specifically, as pointed by \cite{richards2005fundamentals}, the matched filtering (also named as pulse compression) of the signal in \eqref{y_mk} will produce peaks in the range bins, where the targets are located. Thus, we can obtain the coarse range estimations of $G$ targets as $\left\{ r_1,\cdots ,r_g,\cdots ,r_G \right\}$ \cite{xu2015joint,lan2020glrt,xu2023bandwidth,ma2021frac}. $r_g=\mathcal{P} _g\Delta r$ represents the principal range of the $g$th target, where $\mathcal{P} _g$ and $\Delta r=\frac{c}{2B}$ represent the range bin number and bin size, respectively. Note that $B=\Delta f$ denotes the bandwidth of the baseband signal. In other words, the error margin of the true range $R_g$ and the coarse estimate $r_g$ for the $g$th target is within a bin size, i.e., $(R_g-r_g) \in [0,\Delta r] $. 

Then, we estimate the angle of the $g$th target. \textcolor{black}{We construct the joint range-angle spatial spectrum estimation as \cite{xu2015joint,lan2020glrt}
\begin{equation}\label{R_g_Theta_g}
	\begin{aligned}
		&[ \hat{R}_g,\hat{\theta}_g ] =
		\\
		&\mathop {arg\max } \limits_{\begin{array}{c}
				{\theta}_g\in [-90\degree,90\degree] ,\\
				{R}_g\in [r_g-\frac{\Delta r}{2},r_g+\frac{\Delta r}{2}]\\
		\end{array}}\frac{1}{\left| \mathbf{a}_{TR}^{\dagger}({R}_g,{\theta}_g) \mathbf{Q}^{-1}\mathbf{a}_{TR}({R}_g,{\theta}_g) \right|},
	\end{aligned}
\end{equation}
where $\hat{R}_g$ and $\hat{\theta}_g$ denote the estimations of $R_g$ and $\theta _g$. Note that when there are multiple targets in one range bin, targets' angle and distance estimations are obtained by searching for the peaks of \eqref{R_g_Theta_g}.} $\mathbf{Q}=\frac{1}{K}\sum_{k=1}^K{\tilde{\mathbf{y}}_{}^{k}(\tilde{\mathbf{y}}^{k}) ^{\dagger}}$ denotes the sampling covariance matrix of the received signal within one CPI \cite{xu2015deceptive,lan2020suppression},
\begin{equation}\label{a_TR}
	\begin{aligned}
		\mathbf{a}_{TR}(\hat{R}_g,\hat{\theta}_g) =\mathbf{a}_R(\hat{\theta}_g) \otimes [ \mathbf{a}_T(\hat{R}_g) \odot \mathbf{a}_T(\hat{\theta}_g) ] ,
	\end{aligned}
\end{equation}
denotes the transmit-receive steering vector. $\mathbf{a}_R(\hat{\theta}_g)$, $\mathbf{a}_T(\hat{R}_g) $, $\mathbf{a}_T(\hat{\theta}_g) $ are calculated by \eqref{a_T_R_g}, \eqref{a_T_theta}, \eqref{a_R_theta}, respectively.

To estimate targets velocities, the received date of $K$ PRI are stacked into a matrix, as
\begin{equation}\label{y_bmf}
	\begin{aligned}
		\mathbf{Y}&=[\mathbf{y}^{1},\cdots ,\mathbf{y}^{k},\cdots ,\mathbf{y}^{K}]
		\\
		&=\mathbf{AD}+\mathbf{N},
	\end{aligned}
\end{equation}
where $\mathbf{A}=\left[ \mathbf{a}_{TR}^{}(R_1,\theta _1) ,\mathbf{a}_{TR}^{}(R_2,\theta _2) ,\cdots ,\mathbf{a}_{TR}(R_G,\theta _G) \right] \in \mathbb{C} ^{MN\times G}$ denotes the targets manifold matrix. \textcolor{black}{$\mathbf{D}=[ \boldsymbol{\psi }_1(\mathcal{F} _1) ,\boldsymbol{\psi }_2(\mathcal{F} _2) ,\cdots ,\boldsymbol{\psi }_G(\mathcal{F} _G) ] ^T\in \mathbb{C} ^{G\times K}$ is the targets doppler matrix with $\boldsymbol{\psi }_g(\mathcal{F} _g) =[ e^{j2\pi \psi _{g}^{1}},e^{j2\pi \psi _{g}^{2}},...,e^{j2\pi \psi _{g}^{K}} ] ^T$.} $\mathbf{N}=[ \mathbf{n}_{1},\mathbf{n}_{2},\cdots ,\mathbf{n}_{K} ] \in \mathbb{C} ^{MN\times K}$ represents the receive noise matrix over $K$ snapshots.

Then, the LS method estimation is employed to estimate velocities. With the angle and range estimations in \eqref{R_g_Theta_g}, we can write the estimated manifold matrix as $\hat{\mathbf{A}}=[ \mathbf{a}_{TR}( \hat{R}_1,\hat{\theta}_1 ) ,\mathbf{a}_{TR}( \hat{R}_2,\hat{\theta}_2 ) ,\cdots ,\mathbf{a}_{TR}( \hat{R}_G,\hat{\theta}_G ) ]$. \textcolor{black}{The targets doppler matrix is estimated as}
\begin{equation}\label{D_hat}
	\begin{aligned}
		\hat{\mathbf{D}}=\underset{\mathbf{D}}{\min}\left\| \mathbf{Y}-\hat{\mathbf{A}}\mathbf{D} \right\| ^2.
	\end{aligned}
\end{equation}
\textcolor{black}{Solving \eqref{D_hat} yields $\hat{\mathbf{D}}=( \hat{\mathbf{A}}^H\hat{\mathbf{A}} ) ^{-1}\hat{\mathbf{A}}^H\mathbf{Y}$} \footnote{\textcolor{black}{When $M$ or $N$ is large, the $( \hat{\mathbf{A}}^H\hat{\mathbf{A}} ) ^{-1}$ term becomes computationally intensive. Algorithms like blocked matrix inverse or recursive least squares \cite{engel2004kernel} can help mitigate this.}}. Let $\hat{\mathbf{D}}_F$ be the first $K-1$ columns of $\hat{\mathbf{D}}$ and $\hat{\mathbf{D}}_B$ be the second to $K$th columns of $\hat{\mathbf{D}}$. \textcolor{black}{One can observe that the doppler phase matrix is a Vandermont matrix.} Therefore, there exists a rotation vector $\mathbf{d}(\hat{\mathcal{F}}) =[ e^{j2\pi \hat{\mathcal{F}}_1T},\cdots ,e^{j2\pi \hat{\mathcal{F}}_gT},\cdots ,e^{j2\pi \hat{\mathcal{F}}_GT} ] ^T\in \mathbb{C} ^{G\times 1}$ satisfying
\begin{equation}
	\begin{aligned}
		{\hat{\mathbf{D}}_B}^T={\hat{\mathbf{D}}_{F}}^T\mathbf{E}_{\hat{\mathcal{F}}},
	\end{aligned}
\end{equation}
where $\mathbf{E}_{\hat{\mathcal{F}}}=\mathrm{diag}[ \mathbf{d}( \hat{\mathcal{F}} ) ] $. Then, $\mathbf{E}_{\hat{\mathcal{F}}}$ is calculated as $\mathbf{E}_{\hat{\mathcal{F}}}=( \hat{\mathbf{D}}_{F}{\hat{\mathbf{D}}_{F}}^T ) ^{-1}\hat{\mathbf{D}}_{F}{\hat{\mathbf{D}}_B}^T$. 

Finally, the velocity of the $g$th target is estimated as
\begin{equation}\label{F_g}
	\begin{aligned}
		\hat{v}_g=\frac{c\cdot \mathrm{angle} (\kappa _{g,g})}{4f_c\pi T},
	\end{aligned}
\end{equation}
where $\kappa _{g,g}$ denotes the $g$th diagonal element of $\mathbf{E}_{\hat{\mathcal{F}}}$. $\mathrm{angle}\left( \cdot \right) $ means the phase-taking operation.

\subsection{Low-complexity Spatial Spectrum Estimation}
Inspecting \eqref{R_g_Theta_g} reveals that the SSMTE method requires the joint search of 2-D spatial spectrum, which suffers from high complexity. To address this problem, we propose the LCSSE algorithm, which estimates the target angle, distance and velocity by the three-time 1-D search, respectively. Specifically, decomposing the denominator term of \eqref{R_g_Theta_g} yields
\begin{equation}
	\begin{aligned}
		\varOmega (R_g,\theta _g) &=\mathbf{a}_{TR}^{\dagger}(R_g,\theta _g) \mathbf{Q}^{-1}\mathbf{a}_{TR}\left( R_g,\theta _g \right) 
		\\
		&=\mathbf{a}_{T}^{\dagger}(R_g) \mathbf{Z}(\theta_g ) \mathbf{a}_T(R_g), 
	\end{aligned}
\end{equation}
where $\mathbf{Z}(\theta_g) =[ \mathbf{a}_R(\theta _g) \otimes \mathrm{diag}(\mathbf{a}_T(\theta _g))]^{\dagger}\mathbf{Q}^{-1}[ \mathbf{a}_R(\theta _g) \otimes \mathrm{diag}(\mathbf{a}_T(\theta _g)) ] $. Then, $\mathbf{Z}(\theta_g)$ is chunked as 
\begin{equation}
	\begin{aligned}
		\mathbf{Z}(\theta _g) =\left[ \begin{matrix}
			z_1(\theta _g)&		\mathbf{z}_2(\theta _g)\\
			\mathbf{z}_3(\theta _g)&		\mathbf{z}_4(\theta _g)\\
		\end{matrix} \right] , 
	\end{aligned}
\end{equation}
where $z_1(\theta _g) =z_{1,1}(\theta _g) $, $\mathbf{z}_2(\theta _g) =[ z_{1,2}(\theta _g) ,z_{1,3}(\theta _g) , \cdots ,z_{1,N}(\theta _g) ] $, $\mathbf{z}_3(\theta _g) =\mathbf{z}_2(\theta _g) ^{\dagger}$, {\footnotesize $\mathbf{z}_4(\theta _g) =\left[ \begin{matrix}
	z_{2,2}(\theta _g)&		\cdots&		z_{2,N}(\theta _g)\\
	\vdots&		\ddots&		\vdots\\
	z_{N,2}(\theta _g)&		\cdots&		z_{N,N}(\theta _g)\\
\end{matrix} \right] $}. 

Let $\mathbf{a}_T(R_g) =[1,\mathring{\mathbf{a}}_T(R_g) ^T] ^T$ with $\mathring{\mathbf{a}}_T(R_g)=[ e^{j\varphi _2(R_g)},e^{j\varphi _3(R_g)},\cdots ,e^{j\varphi _N(R_g)} ] ^T$. The cost function $\varOmega \left( R_g,\theta _g \right) $ is formulated as
\begin{equation}\label{Ome_the_R}
	\begin{aligned}
		\varOmega (R_g,\theta _g) =&z_1(\theta_g) +\mathring{\mathbf{a}}_{T}^{\dagger}(R_g) \mathbf{z}_{2}^{\dagger}(\theta_g) +\mathbf{z}_2(\theta_g) \mathring{\mathbf{a}}_T(R_g)
		\\
		 &+\mathring{\mathbf{a}}_{T}^{\dagger}(R_g) \mathbf{z}_4(\theta_g) \mathring{\mathbf{a}}_T(R_g) . 
	\end{aligned}
\end{equation}

The partial derivative of $\varOmega (R_g,\theta _g) $ with respect to $\mathring{\mathbf{a}}_T(R_g) $ yields
\begin{equation}\label{weifen1}
	\begin{aligned}
		\frac{\partial \varOmega (R_g,\theta _g)}{\partial \mathring{\mathbf{a}}_T(R_g)}=2\mathbf{z}_{2}^{\dagger}(\theta _g) +2\mathbf{z}_4(\theta _g) \mathring{\mathbf{a}}_T(R_g) . 
	\end{aligned}
\end{equation}

Let \eqref{weifen1} equals to 0, we have $\mathring{\mathbf{a}}_T( R_g ) =-\mathbf{z}_4(\theta _g) ^{-1}\mathbf{z}_{2}^{\dagger}(\theta _g) $. Applying $\mathring{\mathbf{a}}_T( R_g )$ to \eqref{Ome_the_R}, the angle $\theta _g$ is then estimated by
\begin{equation}\label{hat_the_g}
	\begin{aligned}
		[\hat{\theta}_g] =\mathop {arg\max} \limits_{\theta _g\in [-90\degree,90\degree]}\frac{1}{\left|z_1(\theta _g) -\mathbf{z}_{2}(\theta _g) \mathbf{z}_4(\theta _g) ^{-1}\mathbf{z}_{2}^{\dagger}(\theta _g)\right|}. 
	\end{aligned}
\end{equation}
The targets' angle estimations are obtained by searching for the peaks of \eqref{hat_the_g}, named by $\hat{\boldsymbol{\theta}}:({ \hat{\theta}_1,\cdots ,\hat{\theta}_g,\cdots ,\hat{\theta}_G } )$.

Although the targets' angle parameters have now been obtained through \eqref{hat_the_g}, however, their corresponding distances are unknown. To handle this, the angle estimations are brought into \eqref{R_g_Theta_g} one by one. Thereafter, the distance corresponding to $\hat{\theta}_g$ is estimated as
\begin{equation}\label{hat_G_lc}
	\begin{aligned}
		[\hat{R}_g] =\mathop {arg\max} \limits_{R_g\in \boldsymbol{A}_r}\frac{1}{\left| \mathbf{a}_{TR}^{\dagger}(R_g,\hat{\theta}_g) \mathbf{Q}^{-1}\mathbf{a}_{TR}(R_g,\hat{\theta}_g) \right|} ,
	\end{aligned}
\end{equation}
where $\boldsymbol{A}_r=[r_g-\frac{\Delta r}{2},r_g+\frac{\Delta r}{2}] \cup \cdots \cup [r_G-\frac{\Delta r}{2},r_G+\frac{\Delta r}{2}] $ denotes denotes the distance search area.

In summary, in the proposed LCSSE algorithm, the paired angle and distance estimations of the targets are obtained by \eqref{hat_the_g} and \eqref{hat_G_lc}, which are recorded as $(\hat{R}_1,\hat{\theta}_1) ,\cdots , (\hat{R}_G,\hat{\theta}_G) $. Finally, the velocity estimations for each paired target are calculated by \eqref{F_g}.

\subsection{Frequency Offset Design Criterion for Resistance to Range Estimation Ambiguity}
\label{FODCRREA}
Recalling back to \eqref{a_T_R_g}, an unexpected observation is that the transmit range steering vector is a periodic function of distance with period $\frac{c}{2\Delta f}$, i.e., $\mathbf{a}_T(R_g) =\mathbf{a}_T(R_g+\mathfrak{i} \frac{c}{2\Delta f})$, where $\mathfrak{i}$ denotes a positive integer. This characteristic can lead to errors in the pairing of targets angles and ranges. For example, if the positions of target 1 and target 2 are $(R_1,\theta _1)$ and $(R_1+\frac{c}{2\Delta f},\theta _2)$, respectively. Then, the range of target 2 may be estimated to be $R_1$ by \eqref{R_g_Theta_g}. Namely, the target 1 may be misestimated as $(R_1,\theta _2)$.

To address this problem, an effective way is to design the frequency offset of each transmit antenna, so that the phase does not flip periodically over the desired range. Take this into mind, we propose the frequency offset design criterion for resistance to range estimation ambiguity.

Specifically, we design the frequency offset of the $n$th antenna as $\Delta f_n=\varepsilon _n\Delta f$. Note that to guarantee the orthogonality between the transmitted signals, the following condition $\Delta f_{n+1}-\Delta f_n\geqslant \Delta f$ should be satisfied. We split $\varepsilon _n$ as
\begin{equation}\label{varep_n}
	\begin{aligned}
		\varepsilon _n=\mathfrak{i} _n+\mathfrak{q} _n, 
	\end{aligned}
\end{equation}
where $\mathfrak{i} _n$ and $\mathfrak{q} _n$ denote the integer and fractional parts of $\varepsilon _n$, respectively. Then, the $n$th element in $\mathbf{a}_T(R_g) $ is rewritten as 
\begin{equation}
	\begin{aligned}
		e^{-j2\pi \Delta f_n\frac{2R_g}{c}}=e^{-j2\pi \mathfrak{i} _n\Delta f\frac{2R_g}{c}}e^{-j2\pi \mathfrak{q} _n\Delta f\frac{2R_g}{c}}.
	\end{aligned}
\end{equation}
For the $e^{-j2\pi \mathfrak{i} _n\Delta f\frac{2R_g}{c}}$ term, the distance period is $r_{\mathfrak{i} _n}=\frac{c}{2\Delta f}$. On the other hand, the distance period of the $e^{-j2\pi \mathfrak{q} _n\Delta f\frac{2R_g}{c}}$ term is $r_{\mathfrak{q} _n}=\frac{c}{2\Delta f\mathfrak{q} _n}$. Let the distance period of $e^{-j2\pi \Delta f_n\frac{2R_g}{c}}$ be $r_n$, then $r_n$ should be a positive integer multiple of $r_{\mathfrak{i} _n}$ and $r_{\mathfrak{q} _n}$. By guiding of this, we have 
\begin{equation}\label{r_n}
	\begin{aligned}
		r_n=\varpi _nr_{\mathfrak{i} _n}=\varsigma _nr_{\mathfrak{q} _n},
	\end{aligned}
\end{equation}
where $\varpi _n$ and $\varsigma _n$ denote the positive integers to be determined, respectively. In other words, \eqref{r_n} holds at $\frac{\varpi _n}{\varsigma _n}=\frac{1}{\mathfrak{q} _n}$.

Therefore, the period of the transmit range steering vector $\mathbf{a}_T(R_g)$ in the range dimension is equal to the least common multiple (LCM) of the distance periods of all its elements, as
\begin{equation}\label{LCM}
	\begin{aligned}
		r&=\lfloor r_1,\cdots ,r_n,\cdots ,r_N \rfloor _{LCM}
		\\
		&=\frac{c}{2\Delta f}\times \lfloor \varpi _1,\cdots ,\varpi _n,\cdots ,\varpi _N \rfloor _{LCM}.
	\end{aligned}
\end{equation}
In other words, $r$ and $r_n$ should remain positive integer multiples, i.e., $r/r_n=\mathfrak{i}$.

In practice, the maximum system sensing range is $\frac{cT}{2}$. Then the frequency offsets should be designed according to \eqref{varep_n}-\eqref{LCM} such that $r\geqslant \frac{cT}{2}$ to ensure that no distance ambiguity occurs in the interest range.

\subsection{Communication receiver}
Let $\left( \bar{R}_u,\bar{\theta}_u \right)$ be the location of the communication user. Similar to \eqref{S_nk}, the received signal of the $u$th antenna in the $k$th PRI can be expressed as
\begin{equation}\label{y_u_k}
	\begin{aligned}
		y_{u}^{k}\left( t \right) =&\sum_{n=1}^N{h_{u,n}s_{n}^{k}\left( t-\tau _{u,n} \right)}
		\\
		\approx& \sum_{n=1}^N{h_{u,n}\varrho \left( t-kT-\tau _u \right) \tilde{x}_{n}^{k}e^{j2\pi \left( f_c+\Delta f_{n}^{} \right) t}}
		\\
		&\times e^{-j2\pi \Delta f_{n} \frac{2\bar{R}_u}{c}}e^{j2\pi \frac{f_c\left( n-1 \right) d_1\sin \bar{\theta}_u}{c}},
	\end{aligned}
\end{equation}
where $\tau _{u,n}=\frac{\bar{R}_u-(n-1) d_1\sin \bar{\theta}_u-\left( u-1 \right) d_3\sin \bar{\theta}_u}{c}$ and $h_{u,n}\sim \mathcal{C} \mathcal{N} (0,\sigma _{C}^{2}) $ stand for the delay and the channel coefficient between the $n$th transmit antenna and the $u$th receive antenna. $d_3$ represents the adjacent spacing of the receiver. Note that the term $e^{-j2\pi f_c\frac{2\bar{R}_u}{c}}$  is absorbed into the term $h_{u,n}$.

The receiver structure of the communication user is shown in Fig. \ref{receiver_struc}. Similar to the signal demodulation process of the sensing receiver, the output sampled signals of the $n$th channel of the $u$th receive antenna is
\begin{equation}\label{y_u_n_k}
	\begin{aligned}
		\bar{y}_{u,n}^{k}&= y_{u}^{k}\left( t \right) e^{-j2\pi f_ct}e^{-j2\pi \Delta f_nt}*\varrho(t)  +n_{u,n}
		\\
		&=\tilde{x}_{n}^{k}h_{u,n}+n_{u,n},
	\end{aligned}
\end{equation}
where $n_{u,n}\sim \mathcal{C} \mathcal{N} (0,\sigma _{2}^{2}) $ is the receive noise. Note that the constant term $e^{-j2\pi \left( \Delta f_{n}^{}\frac{2\bar{R}_u}{c}-\frac{f_c(n-1) d_1\sin \bar{\theta}_u}{c} \right)}$ is absorbed into $h_{u,n}$. 

Inspecting \eqref{y_u_n_k} reveals that the baseband signals from the $N$ transmit antennas can be separated at the receiver. Leveraging this property, we can combine all demodulated outputs from the same transmit antenna to improve the system BER performance. Specifically, we stack the outputs of the $n$th channels of $U$ receive antennas into a vector as
\begin{equation}\label{y_n_k}
	\begin{aligned}
		\bar{\mathbf{y}}_{n}^{k}&=\left[ \bar{y}_{1,n}^{k},\cdots ,\bar{y}_{u,n}^{k},\cdots ,\bar{y}_{U,n}^{k} \right] 
		\\
		&=\boldsymbol{c}_{}^{T}\boldsymbol{\varsigma }_{n}^{k}\mathbf{h}_{n}^{}+\mathbf{n}_{u}^{},
	\end{aligned}
\end{equation}
where $\mathbf{h}_{n}^{}=[h_{1,n},\cdots ,h_{u,n},\cdots ,h_{U,n}] ^T$ stands for the channel vector between the $n$th transmit antenna and the receiver. $\mathbf{n}_{u}^{}=\left[ n_{1,n},\cdots ,n_{u,n},\cdots ,n_{U,n} \right] ^T$ represents the receive noise vector.

Finally, the maximum likelihood decoder is used to estimate the index and constellation bits emitted by the $n$th antenna as
\begin{equation}\label{y_n_k_bf}
	\begin{aligned}
		\left[ \hat{i}_{n,k},\hat{x}_{n}^{k} \right] =arg\underset{i_{n,k},x_{n}^{k}}{\min}\left\| \bar{\mathbf{y}}_{n}^{k}-c_{i_{n,k}}^{}x_{n}^{k}\mathbf{h}_{n}^{} \right\| ^2,
	\end{aligned}
\end{equation}
where $\hat{i}_{n,k}$, $\hat{x}_{n}^{k}$ denote the estimations of $i_{n,k}$, $x_{n}^{k}$. Note that the transmitted index and constellation bits from all transmit antenna are sequentially estimated by \eqref{y_n_k} and \eqref{y_n_k_bf}. 

\section{System Performace Analysis}
\label{SPA}
In this paper, the widely used BER and CRB metrics are considered to evaluate the system communication and sensing performance, respectively. In this section, closed expressions for the system BER and CRB are derived. Moreover, the system sensing complexity is analyzed.

\subsection{System CRB Analysis}
Within one CPI, the noiseless data matrix for a target located at $(\theta _g,R_g)$ can be rewritten as
\begin{equation}\label{pi_symbol}
	\begin{aligned}
		\mathbf{\Pi }&=\xi _g\left\{ \mathbf{a}_R(\theta _g) \otimes \left[ \mathbf{a}_T(R_g) \odot \mathbf{a}_T(\theta _g) \right] \right\} \boldsymbol{\psi }_{g}^{T}(\mathcal{F} _g)
		\\
		&=\xi _g\mathbf{W},
	\end{aligned}
\end{equation}
where $\mathbf{W}=\mathbf{a}_{TR}(\theta _g,R_g) \boldsymbol{\psi }_{g}^{T}(\mathcal{F} _g)$ and $\mathbf{a}_{TR}(\theta _g,R_g)$ is denoted in \eqref{a_TR}.

For convenience, define the unknown parameter vector as
\begin{equation} \label{rho}
	\begin{aligned}
		\boldsymbol{\rho }= [\mathrm{Re}\left\{ \xi _g \right\} ,\mathrm{Im}\left\{ \xi _g \right\} ,R_g,\theta _g,\mathcal{F} _g] ^T
		.
	\end{aligned}
\end{equation}

According to the CRB definition, the estimation accuracy lower bound of $\boldsymbol{\rho }$ is given by the diagonal elements of $\mathbf{F}^{-1}$. $\mathbf{F}\in \mathbb{C} ^{5\times 5}$ represents the Fisher information matrix of the received signal, whose $(x,y)$th element is given by \cite{xu2015deceptive}
\begin{equation} \label{F_x_y}
	\begin{aligned}
		F_{x,y}&=2\mathrm{Re}\left\{ \mathrm{Tr}\left[ \frac{\partial \mathbf{\Pi }^{\dagger}}{\partial \rho _x}\mathbf{\Lambda }^{-1}\frac{\partial \mathbf{\Pi }}{\partial \rho _y} \right] \right\} 
		\\
		&=2\mathrm{Re}\left\{ \mathrm{Tr}\left\{ \frac{\partial (\xi _g\mathbf{W}) ^{\dagger}}{\partial \rho _x}\mathbf{\Lambda }^{-1}\frac{\partial (\xi _g\mathbf{W})}{\partial \rho _y} \right\} \right\} 
		,
	\end{aligned}
\end{equation}
where $\mathbf{\Lambda }=\sigma _{1}^{2}\boldsymbol{I}_{N\times M}$ represents the noise covariance matrix. $\rho _{x/y}$ denotes the $x/y$th element of $\boldsymbol{\rho }$. Substituting \eqref{a_TR}, \eqref{rho} into \eqref{F_x_y}, we have
\begin{equation}  
	\begin{aligned}
		\frac{\partial (\xi _g\mathbf{W})}{\partial \mathrm{Re}\left\{ \xi _g \right\}}=\mathbf{W}
		,
	\end{aligned}
\end{equation}
\begin{equation}  
	\begin{aligned}
		\frac{\partial (\xi _g\mathbf{W})}{\partial \mathrm{Im}\left\{ \xi _g \right\}}=j\mathbf{W}
		,
	\end{aligned}
\end{equation}
\begin{equation}  
	\begin{aligned}
		\frac{\partial(\xi _g\mathbf{W})}{\partial R_g}=\xi _g\left( \mathbf{a}_R(\theta _g) \otimes \left[ \dot{\mathbf{a}}_T(R_g) \odot \mathbf{a}_T(\theta _g)\right] \right) \boldsymbol{\psi }_{g}^{T}(\mathcal{F} _g)
		,
	\end{aligned}
\end{equation}
\begin{equation}  
	\begin{aligned}
		\frac{\partial (\xi _g\mathbf{W})}{\partial \theta _g}
		=&\xi _g\left\{ \dot{\mathbf{a}}_R(\theta _g) \otimes \left[ \mathbf{a}_T(R_g) \odot \mathbf{a}_T(\theta _g) \right] \right. \boldsymbol{\psi }_{g}^{T}(\mathcal{F} _g)
		\\
		&+\left. \mathbf{a}_R(\theta _g) \otimes \left[ \mathbf{a}_T(R_g) \odot \dot{\mathbf{a}}_T (\theta _g) \right] \right\} \boldsymbol{\psi }_{g}^{T}(\mathcal{F} _g)
		,
	\end{aligned}
\end{equation}
\begin{equation}  
	\begin{aligned}
		\frac{\partial (\xi _g\mathbf{W})}{\partial \mathcal{F} _g}=\xi _g\mathbf{a}_R(\theta _g) \otimes \left[ \mathbf{a}_T(R_g) \odot \mathbf{a}_T(\theta _g) \right] \dot{\boldsymbol{\psi}}_{g}^{T}(\mathcal{F} _g) 
		,
	\end{aligned}
\end{equation}
where $\dot{\mathbf{a}}_T(R_g) =\mathbf{E}_{T,R}\mathbf{a}_T(R_g)$, $\dot{\mathbf{a}}_R(\theta _g) =\mathbf{E}_{R,\theta}\mathbf{a}_R(\theta _g)$, $\dot{\mathbf{a}}_T(\theta _g) =\mathbf{E}_{T,\theta}\mathbf{a}_T(\theta _g) $ and $\dot{\boldsymbol{\psi}}_{g} (\mathcal{F} _g) =\mathbf{E}_{\mathcal{F}}\boldsymbol{\psi }_{g}(\mathcal{F} _g)$ with
\begin{equation}  
	\begin{aligned}
		\mathbf{E}_{T,R}=-j\frac{4\pi}{c}\Delta f\mathrm{diag}\left\{ 0,1,\cdots ,N-1 \right\} 
		,
	\end{aligned}
\end{equation}
\begin{equation}  
	\begin{aligned}
		\mathbf{E}_{R,\theta}=j\frac{2\pi f_cd_2}{c}\cos \theta _g\mathrm{diag}\left\{ 0,1,\cdots ,M-1 \right\} 
		,
	\end{aligned}
\end{equation}
\begin{equation}  
	\begin{aligned}
		\mathbf{E}_{T,\theta}=j\frac{2\pi f_cd_1}{c}\cos \theta _g\mathrm{diag}\left\{ 0,1,\cdots ,N-1 \right\} 
		,
	\end{aligned}
\end{equation}
\begin{equation}  
	\begin{aligned}
		\mathbf{E}_{\mathcal{F}}=j2\pi T\mathrm{diag}\left\{ 0,1,\cdots ,K-1 \right\} 
		.
	\end{aligned}
\end{equation}

Then, the Fisher information matrix can be rewritten as \eqref{F_Mat},
\begin{figure*}[hbpt]
	\begin{equation}\label{F_Mat}
		\begin{aligned}
			\mathbf{F}=\frac{2P_{\mathrm{S}}}{N}\mathrm{Re}\left\{ \left[ \begin{matrix}
				\mathrm{Tr}\left\{ \zeta ^{\dagger}\zeta \right\}&		0&		\mathrm{Tr}\left\{ \xi _g\zeta ^{\dagger}\zeta _R \right\}&		\mathrm{Tr}\left\{ \xi _g\zeta ^{\dagger}\zeta _{\theta} \right\}&		\mathrm{Tr}\left\{ \xi _g\zeta ^{\dagger}\zeta _{\mathcal{F}} \right\}\\
				0&		\mathrm{Tr}\left\{ \zeta ^{\dagger}\zeta \right\}&		\mathrm{Tr}\left\{ -j\zeta ^{\dagger}\xi _g\zeta _R \right\}&		\mathrm{Tr}\left\{ -j\xi _g\zeta ^{\dagger}\zeta _{\theta} \right\}&		\mathrm{Tr}\left\{ -j\xi _g\zeta ^{\dagger}\zeta _{\mathcal{F}} \right\}\\
				\mathrm{Tr}\left\{ {\xi _g}^{\dagger}{\zeta _R}^{\dagger}\zeta \right\}&		\mathrm{Tr}\left\{ {j\xi _g}^{\dagger}{\zeta _R}^{\dagger}\zeta \right\}&		\mathrm{Tr}\left\{ \left| \xi _g \right|^2{\zeta _R}^{\dagger}\zeta _R \right\}&		\mathrm{Tr}\left\{ \left| \xi _g \right|^2{\zeta _R}^{\dagger}\zeta _{\theta} \right\}&		\mathrm{Tr}\left\{ \left| \xi _g \right|^2{\zeta _R}^{\dagger}\zeta _{\mathcal{F}} \right\}\\
				\mathrm{Tr}\left\{ {\xi _g}^{\dagger}{\zeta _{\theta}}^{\dagger}\zeta \right\}&		\mathrm{Tr}\left\{ {j\xi _g}^{\dagger}{\zeta _{\theta}}^{\dagger}\zeta \right\}&		\mathrm{Tr}\left\{ \left| \xi _g \right|^2{\zeta _{\theta}}^{\dagger}\zeta _R \right\}&		\mathrm{Tr}\left\{ \left| \xi _g \right|^2{\zeta _{\theta}}^{\dagger}\zeta _{\theta} \right\}&		\mathrm{Tr}\left\{ \left| \xi _g \right|^2{\zeta _{\theta}}^{\dagger}\zeta _{\mathcal{F}} \right\}\\
				\mathrm{Tr}\left\{ {\xi _g}^{\dagger}{\zeta _{\mathcal{F}}}^{\dagger}\zeta \right\}&		\mathrm{Tr}\left\{ {j\xi _g}^{\dagger}{\zeta _{\mathcal{F}}}^{\dagger}\zeta \right\}&		\mathrm{Tr}\left\{ \left| \xi _g \right|^2{\zeta _{\mathcal{F}}}^{\dagger}\zeta _R \right\}&		\mathrm{Tr}\left\{ \left| \xi _g \right|^2{\zeta _{\mathcal{F}}}^{\dagger}\zeta _{\theta} \right\}&		\mathrm{Tr}\left\{ \left| \xi _g \right|^2{\zeta _{\mathcal{F}}}^{\dagger}\zeta _{\mathcal{F}} \right\}\\
			\end{matrix} \right] \right\},
		\end{aligned}
	\end{equation}
\end{figure*}
where $\zeta _R=\mathbf{\Lambda }^{-\frac{1}{2}}\frac{\partial \mathbf{W}}{\partial R_g}$, $\zeta _{\theta}=\mathbf{\Lambda }^{-\frac{1}{2}}\frac{\partial \mathbf{W}}{\partial \theta _g}$, $\zeta _{\mathcal{F}}=\mathbf{\Lambda }^{-\frac{1}{2}}\frac{\partial \mathbf{W}}{\partial \mathcal{F} _g}$ and $\zeta =\mathbf{\Lambda }^{-\frac{1}{2}}\mathbf{W}$.

Further, $\mathbf{F}^{-1}$ is represented by 
\begin{equation}  
	\begin{aligned}
		\mathbf{F}^{-1}=\frac{1}{2}\left[ \begin{matrix}
			\mathbf{F}_{11}&		\mathbf{F}_{12}\\
			\mathbf{F}_{21}&		\mathbf{F}_{22}\\
		\end{matrix} \right] ^{-1}=\frac{1}{2}\left[ \begin{matrix}
			\times&		\times\\
			\times&		\mathbf{D}^{-1}\\
		\end{matrix} \right] ,
	\end{aligned}
\end{equation}
where $\mathbf{F}_{11}\in \mathbb{C} ^{2\times 2}$, $\mathbf{F}_{12}\in \mathbb{C} ^{2\times 3}$, $\mathbf{F}_{21}\in \mathbb{C} ^{3\times 2}$ and $\mathbf{F}_{22}\in \mathbb{C} ^{3\times 3}$ are the chunking matrices in $\mathbf{F}$. Note that the diagonal elements of $\mathbf{D}^{-1}$ contain the  estimation information for angle, distance, and velocity. According to the chunked matrix inverse formula \cite{xu2015deceptive}, $\mathbf{D}$ is calculated as
\begin{equation}  
	\begin{aligned}
		\mathbf{D}=\mathbf{F}_{22}-\mathbf{F}_{21}{\mathbf{F}_{11}}^{-1}\mathbf{F}_{12}=\left[ \begin{matrix}
			D_{11}&		D_{12}&		D_{13}\\
			D_{21}&		D_{22}&		D_{23}\\
			D_{31}&		D_{32}&		D_{33}\\
		\end{matrix} \right] ,
	\end{aligned}
\end{equation}
where $\mathbf{D}_{i,j}$ denotes the $i,j$th element in $\mathbf{D}$.

Finally, the CRBs of angle, distance, doppler frequency estimations are given by
\begin{equation}  
	\begin{aligned}
		\mathrm{CRB}_{R_g}=\frac{1}{2\det \left( \mathbf{D} \right)}\det \left( \left[ \begin{matrix}
			D_{22}&		D_{23}\\
			D_{32}&		D_{33}\\
		\end{matrix} \right] \right),
	\end{aligned}
\end{equation}
\begin{equation}  
	\begin{aligned}
		\mathrm{CRB}_{\theta _g}=\frac{1}{2\det \left( \mathbf{D} \right)}\det \left( \left[ \begin{matrix}
			D_{11}&		D_{13}\\
			D_{31}&		D_{33}\\
		\end{matrix} \right] \right) ,
	\end{aligned}
\end{equation}
and
\begin{equation}  
	\begin{aligned}
		\mathrm{CRB}_{\mathcal{F}}=\frac{1}{2\det \left( \mathbf{D} \right)}\det \left( \left[ \begin{matrix}
			D_{11}&		D_{12}\\
			D_{21}&		D_{22}\\
		\end{matrix} \right] \right) ,
	\end{aligned}
\end{equation}
respectively, where $\det (\cdot) $ denotes the determinant operator.

\subsection{Complexity Analysis of System Sensing Methods}
We analyze the complexity of the proposed SSMTE and LCSSE algorithms by counting the required multiplication operations. The complexity of computing $\mathbf{Q}^{-1}$ is $\mathcal{O} \left\{ K(NM) ^2+(NM) ^3 \right\}$. For mathematical convenience, let $s_{r}$ and $s_{\theta}$ denote the number of distance and angle search steps within a range bin, respectively. Then the angle-distance estimation complexity in \eqref{R_g_Theta_g} is $\mathcal{O} \left\{s_rs_{\theta}( N^2M^2+NM+1 )\right\}$. For velocity estimation, the computational complexity of $\hat{\mathbf{D}}$ and $\mathbf{E}_{\hat{\mathcal{F}}}$ are $\mathcal{O} \left\{ G^3+2G^2NM+GMNK \right\} $ and $\mathcal{O} \left\{ 2G^3+2G^2 (K-1)  \right\} $, respectively. The complexity of computing $G$ target speeds in \eqref{F_g} is $\mathcal{O} \left\{ 4G \right\} $. Assuming that there are $G'\leqslant G$ targets located at different range bins, the complexity of the SSMTE method is calculated by
\begin{equation}  
	\begin{aligned}
		\mathcal{O} \left\{ \begin{array}{c}
			KN^2M^2+N^3M^3+s_rs_{\theta}G'(N^2M^2+NM+1)\\
			+3G^3+2G^2(K+NM-1) +GMNK+4G\\
		\end{array} \right\}  .
	\end{aligned}
\end{equation}

For the LCSSE method, the computational complexity of $\mathbf{Z}(\theta _g) $ is $\mathcal{O} \left\{ N^2M^2+NM \right\} $. Then, The complexity of \eqref{hat_the_g} is $\mathcal{O} \left\{ s_{\theta}[(N-1) ^3+(N-1) ^2+(N-1) +1] \right\} $, while \eqref{hat_G_lc} costs $\mathcal{O} \left\{ s_rG' (N^2M^2+NM+1) \right\} $. Finally, the complexity of the LCSSE method is computed as
\begin{equation}  
	\begin{aligned}
		\mathcal{O} \left\{ \begin{array}{c}
			KN^2M^2+N^3M^3+s_{\theta} [(N-1) ^3+(N-1) ^2+N]\\
			+s_rGG' (N^2M^2+NM+1) \\
			+3G^3+2G^2(K+NM-1) +GMNK+4G\\
		\end{array} \right\}.
	\end{aligned}
\end{equation}

\subsection{System BER Upper Bound Analysis}
Referring back to Section \ref{sc2}, we know that at the BS side, the information bits carried by one antenna can be categorized into index bits $\mu _I=\log _2J$ and constellation bits $\mu _C=\log _2L$. Therefore, the system average bit error rate (ABER) is formulated as
\begin{equation}
	\begin{aligned}
		P_{\mathrm{CCIE}}=\frac{\sum_{n=1}^N{(P_{I,n}\mu _I+P_{C,n}\mu _C)}}{N(\mu _I+\mu _C )}
		,
	\end{aligned}
\end{equation}
where $P_{I,n}$, $P_{C,n}$ denote the average error probability (ABEP) of the index and constellation bits carried by the $n$th transmit antenna, respectively.

We first derive $P_{I,n}$. One can observe that there are $C_{\mu _I}^{e}$ events that $e\in \left[ 1,\mu _I \right]$ bits errors out of $\mu _I$ bits. The misestimated index has the same probability of being the remaining $J-1$ complex coefficients. Thus, the ABEP of the indexed bits can be modeled as
\begin{equation} \label{P_In}
	\begin{aligned}
		P_{I,n}&=\frac{P_{IM}}{(2^{\mu _I}-1) \mu _I}\sum_{e=1}^{\mu _I}{e}C_{\mu _I}^{e}
		\\
		&=\frac{2^{\mu _I}P_{IM}}{2 (2^{\mu _I}-1)}
		,
	\end{aligned}
\end{equation}
where $P_{IM}$ denotes the probability that the selected complex coefficient is incorrectly detected.

The $P_{IM}$ can be derived by the union bounding technique. Specifically, calling back to \eqref{y_n_k_bf}, the conditional pairwise error probability (PEP) that $i_{n,k}$ is erroneously detected as $i_{n,k}^{'}$ on $\mathbf{h}_{n}$ can be formulated as 
\begin{equation}  \label{Pr_PEP}
	\begin{aligned}
		&\mathrm{Pr(}i_{n,k}\rightarrow i_{n,k}^{\prime}\mid \mathbf{h}_{n}^{})
		\\
		&=\mathrm{Pr}\left( \left\| \bar{\mathbf{y}}_{n}^{k}-c_{i_{n,k}}^{}x_{n}^{k}\mathbf{h}_{n}^{} \right\| ^2>\left\| \bar{\mathbf{y}}_{n}^{k}-c_{i_{n,k}^{\prime}}^{}x_{n\prime}^{k}\mathbf{h}_{n}^{} \right\| ^2 \right) 
		\\
		&=Q\left( \sqrt{\frac{\kappa}{2\sigma _{2}^{2}}} \right) 
		,
	\end{aligned}
\end{equation}
where $\kappa =\left\| c_{i_{n,k}^{\prime}}^{}x_{n\prime}^{k}\mathbf{h}_{n} -c_{i_{n,k}} x_{n}^{k}\mathbf{h}_{n} \right\| ^2$.

Since the elements in $\mathbf{h}_{n}$ follow i.i.d. $\mathcal{C} \mathcal{N} (0,\sigma _{C}^{2})$, $\kappa$ in \eqref{Pr_PEP} can be rewritten as $\kappa =\sum_{u=1}^{2U}{\varpi _{u}^{2}}$, where $\varpi _{u}^{2}\sim \mathcal{C} \mathcal{N} (0,\sigma _{\kappa}^{2}) $ with
\begin{equation} \label{kappa_var}
	\begin{aligned}
		\left\{ \begin{array}{c}
			\sigma _{\kappa}^{2}=\frac{\left| c_{i_{n,k}^{'}} x_{n'}^{k}-c_{i_{n,k}} x_{n}^{k} \right|^2\sigma _{C}^{2}}{2},\mathrm{if} i_{n,k}^{'}\ne i_{n,k}^{}\\
			\sigma _{\kappa}^{2}=\frac{\left| x_{n'}^{k}-x_{n}^{k} \right|^2\left| c_{i_{n,k}} \right|^2\sigma _{C}^{2}}{2},\mathrm{if} i_{n,k}^{'}=i_{n,k} .\\
		\end{array} \right 
		.
	\end{aligned}
\end{equation}

From \eqref{kappa_var}, one can observe that $\kappa$ follows the chi-square distribution with $2U$ degrees of freedom (DoF), whose probability density function (PDF) is written as
\begin{equation} \label{f_kappa}
	\begin{aligned}
		f_{\kappa}\left( x \right) =\frac{1}{2^U\Gamma \left( U \right) \left( \sigma _{\kappa}^{2} \right) ^U}x^{U-1}\exp \left( -\frac{x}{2\sigma _{\kappa}^{2}} \right) 
		.
	\end{aligned}
\end{equation}
Averaging \eqref{f_kappa} on $\kappa$ gives that
\begin{equation} 
	\begin{aligned}
		&\mathrm{Pr(}i_{n,k}\rightarrow i_{n,k}^{\prime})=\int_0^{\infty}{Q\left( \sqrt{\frac{\kappa}{2\sigma _{2}^{2}}} \right) f_{\kappa}\left( x \right)}dx
		\\
		&=\left. \left[ P\left( \alpha \right) \right] ^U\sum_{u=0}^{U-1}{\left( \begin{array}{c}
				U-1+u\\
				u\\
			\end{array} \right.} \right) \left[ 1-P\left( \alpha \right) \right] ^u 
		,
	\end{aligned}
\end{equation}
where
\begin{equation} 
	\begin{aligned}
		P(\alpha )=\frac{1}{2}\left( 1-\sqrt{\frac{\alpha}{1+\alpha}} \right) 
		,
	\end{aligned}
\end{equation}
with $\alpha =\frac{\sigma _{\kappa}^{2}}{2\sigma _{2}^{2}}$.

According to the union bound technique, the tight upper bound of $P_{IM}$ can  be expressed as \cite{simon2001digital}
\begin{equation} \label{P_IM}
	\begin{aligned}
		P_{IM}\leqslant \frac{1}{JL}\sum_{\begin{array}{c}
				i_{n,k},i_{n,k}^{\prime}\\
				i_{n,k}\ne i_{n,k}^{\prime}\\
		\end{array}}{\sum_{x_{n}^{k},x_{n\prime}^{k}}{\mathrm{Pr(}i_{n,k}\rightarrow i_{n,k}^{\prime})}}
		.
	\end{aligned}
\end{equation}

Next, we derive the ABEP of the constellation bits, i.e., $P_{C,n}$. We found that $P_{C,n}$ consists of two parts, one is the index bits are correctly detected but the QAM symbol is incorrectly detected. The other part is the case where the index bits are incorrectly detected, resulting in incorrect detection of the QAM symbol. Take this into mind, we have
\begin{equation} \label{P_cn}
	\begin{aligned}
		P_{C,n}=\frac{J-1}{J}P_{IM}+\left( 1-P_{IM} \right) P_{QAM}
		,
	\end{aligned}
\end{equation}
where $P_{QAM}$ denotes the ABEP of the constellation bits when the complex coefficient estimate is wrong, which is derived in the sequel.

We take a $L$-ary QAM symbol that can be split into two pulse amplitude modulation (PAM) symbols: $\upsilon$-ary PAM of the I-signal and $\omega$-ary PAM of the Q-signal, $L=\upsilon \times \omega$. The conditional probability that the $q$th bit errors in the I-signal component can be expressed as
\begin{equation} \label{P_upsilon}
	\begin{aligned}
		&P_{\upsilon}\left( q \middle| \gamma _{j,l} \right) =\frac{2}{\upsilon}\sum_{i=0}^{(1-2^{-q}) \upsilon -1}{\left\{  (-1) ^{\lfloor \frac{i\cdot 2^{q-1}}{\mu} \rfloor} \right.}\\
		&\times \left( 2^{q-1}-\left. \lfloor \frac{i\cdot 2^{q-1}}{\left. \upsilon \right.}+\frac{1}{2} \rfloor \right. \right) \left. Q\left( (2i+1)\varepsilon \sqrt{2\gamma _{j,l} } \right) \right\} 
		,
	\end{aligned}
\end{equation}
where $\varepsilon =\sqrt{\frac{3}{\upsilon ^2+\omega ^2-2}}$ denotes the minimum norm distance between two constellation points. $\gamma _{j,l}^{}=\frac{\left\| c_{j}^{}x_{l}^{}\mathbf{h}_{n}^{} \right\| ^2}{\sigma _{2}^{2}}$ denotes the instantaneous total received signal to noise ratio (SNR) on the $\mathbf{h}_{n}$ when transmit the complex coefficient $c_{j}$ and the constellation symbol of $x_{l}$, the PDF of which is given by \cite{li2019spatial}
\begin{equation} 
	\begin{aligned}
		f_{\gamma _{j,l} }(x)=\frac{1}{2^U\Gamma \left( U \right) \left( \frac{\left| c_{j}x_{l} \right|^2\sigma _{C}^{2}}{2\sigma _{2}^{2}} \right) ^U}x^{U-1}\exp \left( -\frac{\sigma _{2}^{2}}{\left| c_{j}x_{l} \right|^2\sigma _{C}^{2}}x \right)
		.
	\end{aligned}
\end{equation}

By averaging \eqref{P_upsilon} on $\gamma _{j,l}$, we have
\begin{equation} 
	\begin{aligned}
		&P_{\upsilon}^{j,l} (q) \\
		&=\frac{2}{\upsilon}\sum_{i=0}^{ (1-2^{-q}) \upsilon -1}{\left\{ (-1) ^{\lfloor \frac{i\cdot 2^{q-1}}{\upsilon} \rfloor}\left( 2^{q-1}-\left. \lfloor \frac{i\cdot 2^{q-1}}{\left. \upsilon \right.}+\frac{1}{2} \rfloor \right. \right) \right.}
		\\
		&\times \left. \left. \left[ P\left( \alpha _{j,l}' \right) \right] ^U\sum_{u=0}^{U-1}{\left( \begin{array}{c}
				U-1+u\\
				u\\
			\end{array} \right.} \right) \left[ 1-P (\alpha _{j,l}') \right] ^u \right\} 
		,
	\end{aligned}
\end{equation}
where 
\begin{equation} 
	\begin{aligned}
		P\left( \alpha _{j,l}' \right) =\frac{1}{2}\left( 1-\sqrt{\frac{\alpha _{j,l}'}{1+\alpha _{j,l}'}} \right) 
		,
	\end{aligned}
\end{equation}
with $\alpha _{j,l}'=\frac{\left[ (2i+1)\varepsilon \right] ^2\left| c_{j} x_{l}  \right|^2\sigma _{C}^{2}}{\sigma _{2}^{2}}$.

Similarly, the error probability of the $q$th bit in the $\omega$-ary PAM component can be expressed as
\begin{equation} 
	\begin{aligned}
		&P_{\omega}^{j,l}(q) \\
		&=\frac{2}{\omega}\sum_{i=0}^{(1-2^{-q}) \omega -1}{\left\{ (-1) ^{\lfloor \frac{i\cdot 2^{q-1}}{\omega} \rfloor}\left( 2^{q-1}-\left. \lfloor \frac{i\cdot 2^{q-1}}{\omega}+\frac{1}{2} \rfloor \right. \right) \right.}
		\\
		&\times \left. \left[ P (\alpha _{j,l}') \right] ^U\sum_{u=0}^{U-1}{\left( \begin{array}{c}
				U-1+u\\
				u\\
			\end{array} \right)}\left[ 1-P(\alpha _{j,l}') \right] ^u \right\} 
		.
	\end{aligned}
\end{equation}

Therefore, with emission complex coefficients of $c_{i}$, the ABEP of the constellation symbol $x_{n}$ is calculated as
\begin{equation} 
	\begin{aligned}
		P_{j,l} =\frac{1}{\log _2L}\left[ \sum_{q=1}^{\log _2\upsilon}{P_{\upsilon}^{j,l}(q)}+\sum_{q=1}^{\log _2\omega}{P_{\omega}^{j,l}(q)} \right] 
		.
	\end{aligned}
\end{equation}

Further, the ABEP of $P_{QAM}$ is obtained as
\begin{equation} \label{P_QAM}
	\begin{aligned}
		P_{QAM}=\frac{1}{JL}\sum_{j=1}^J{\sum_{l=1}^L{P_{j,l} }}
		.
	\end{aligned}
\end{equation}

Substituting \eqref{P_In}, \eqref{P_IM}, \eqref{P_cn} and \eqref{P_QAM} into \eqref{P_cn}, the system ABER can be expressed as 
\begin{equation} \label{P_CCIE}
	\begin{aligned}
		P_{\mathrm{CCIE}}=\frac{\frac{2^{\mu _I}P_{IM}}{2 (2^{\mu _I}-1)}\mu _I+\left[ \frac{1}{2}P_{IM}+(1-P_{IM}) P_{QAM} \right] \mu _C}{\mu _I+\mu _C}
		.
	\end{aligned}
\end{equation}

\section{Simulation Results}
\label{sc5}
In this section, we perform Monte Carlo simulations to evaluate the proposed ISAC system performance and verify analytical results. \textcolor{black}{The carrier frequency is set as $f_c=10 \mathrm{GHz}$, which is often used in radar systems \cite{bole2013radar}, satellite and terrestrial communications \cite{graf1999modern}.} Unless specified, the main parameters used in the experimental study are set to $d_1=d_2=c/f_c$, $T=60 \mathrm{\mu s}$, $T_W=20 \mathrm{\mu s}$, $\Delta f=2\mathrm{MHz}$. $\sigma _{C}^{2}=1$. The SNR of the sensing receiver and the communication receiver are denoted as $\frac{1}{\sigma _{1}^{2}}$ and $\frac{1}{\sigma _{2}^{2}}$, respectively. 

\subsection{Sensing Simulation}
\begin{figure}[htbp]	
	\centering
	\subfigure[]	
	{\includegraphics[width=0.24\textwidth]{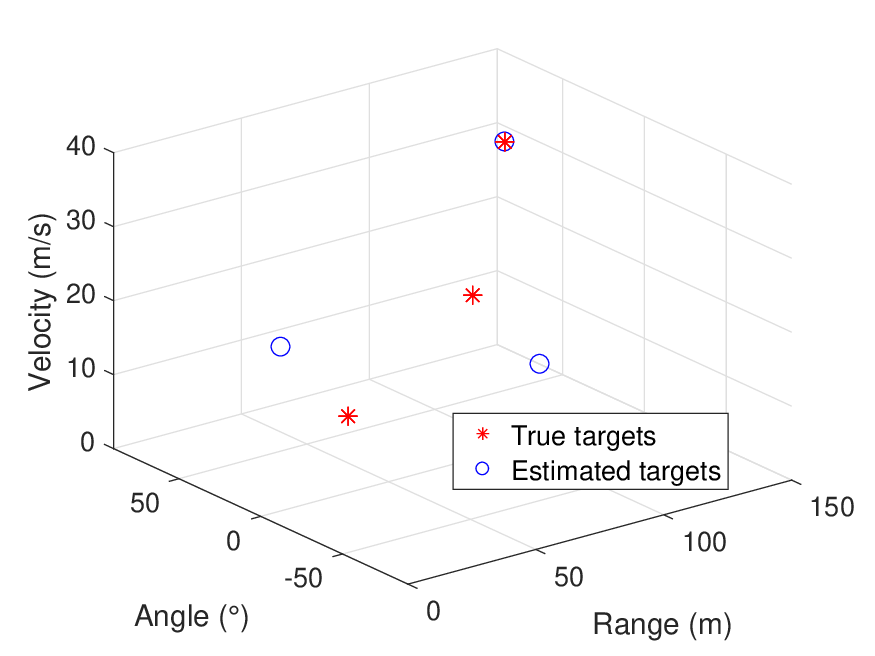}}\label{3D_with_amb}
	\subfigure[]	{\includegraphics[width=0.24\textwidth]{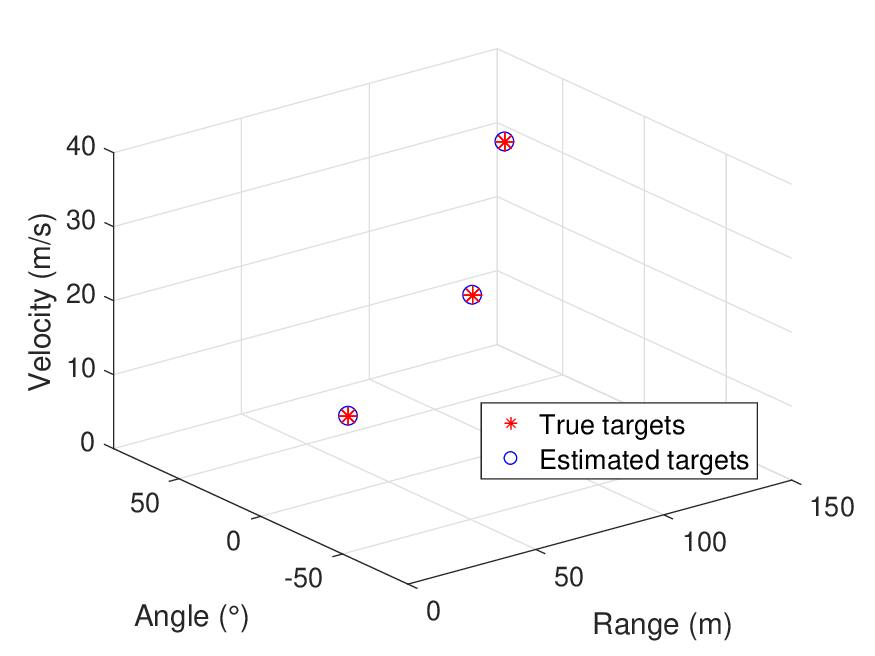}}\label{3D_without_amb}
	\subfigure[]	
	{\includegraphics[width=0.24\textwidth]{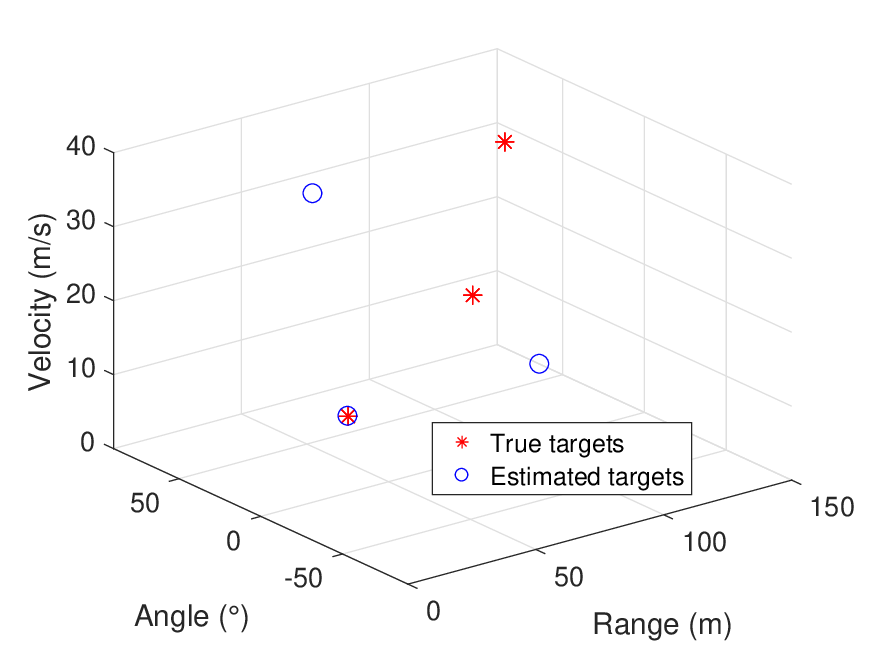}}\label{LCSSE_3D_confu}
	\subfigure[]	
	{\includegraphics[width=0.24\textwidth]{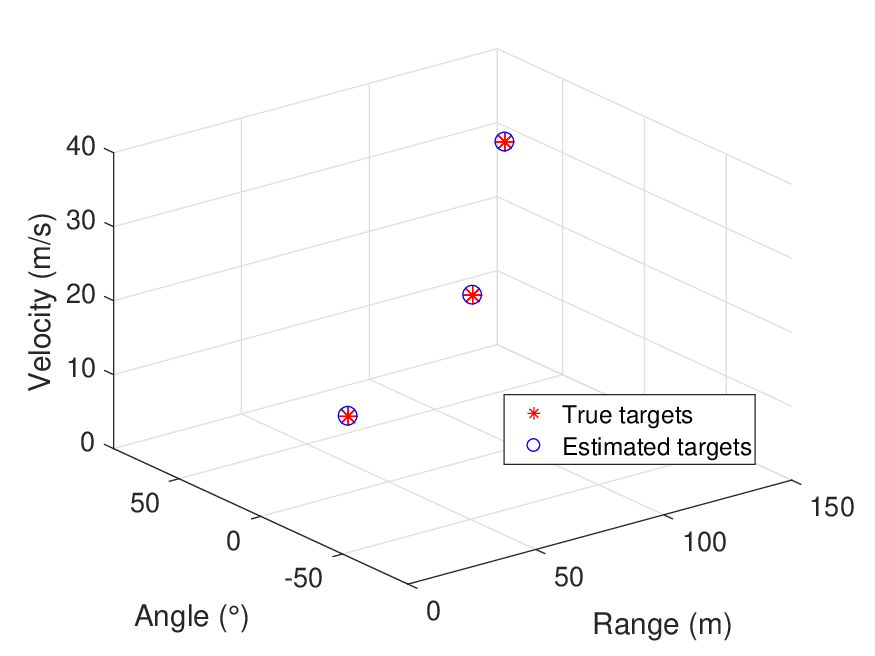}}\label{LCSSE_3D_non_confu}
	\caption{Targets recovery results, where  $N=M=6$, $\mathrm{SNR}=5\mathrm{dB}$, $K=200$. (a) SSMTE method with LFO, (b) SSMTE method with FODC, (c) LCSSE method with LFO, (d) LCSSE method with FODC.}
	\label{3D_amb1}
\end{figure}

In this subsection, the root mean square errors (RMSE) and hit rate are adopted to evaluate the system sensing performance. A hit is proclaimed if the sum of the angle, distance and velocity estimation errors for the three targets in Fig. \ref{3D_amb1} is less than 0.2 \cite{huang2020majorcom}. The RMSE is defined as
\begin{equation} 
	\begin{aligned}
		RMSE=\frac{1}{G}\sum_{g=1}^G{\sqrt{\frac{1}{M}\sum_{m=1}^M{ ( \rho _g-\hat{\rho}_{g,m}) ^2}}},
	\end{aligned}
\end{equation}
where $M$ stands for the number of Monte Carlo trails. $\rho _g$ denotes the true value of $R_g$, $\theta _g$ or $v _g$, while $\hat{\rho}_{g,m}$ representing the estimation of $\rho _g$ in the $m$th trail.  

The frequency offsets design criterion proposed in Section \ref{FODCRREA} and widely used linear frequency offsets \cite{lan2020glrt,xu2015joint,lan2020suppression} are denoted as 'FODC' and 'LFO' in simulations, respectively. Note that the 'FODC' transmit frequency offsets are set as $\left\{\left\{ 0,1,2,3.17,4.2,5.2 \right\}\times \Delta f \right\} \mathrm{MHz}$, while the 'LFO' transmit frequency offsets are set as  $\left\{ \left\{0,1,2,3,4,5\right\}\times \Delta f  \right\} \mathrm{MHz}$. The proposed sensing methods are compared with FDA-MIMO-based frequency offset permutation index modulation (FOPIM) scheme \cite{jian2023fda}. 

\begin{figure}[htbp]
	\centering
	\subfigure[]	
	{\includegraphics[width=0.24\textwidth]{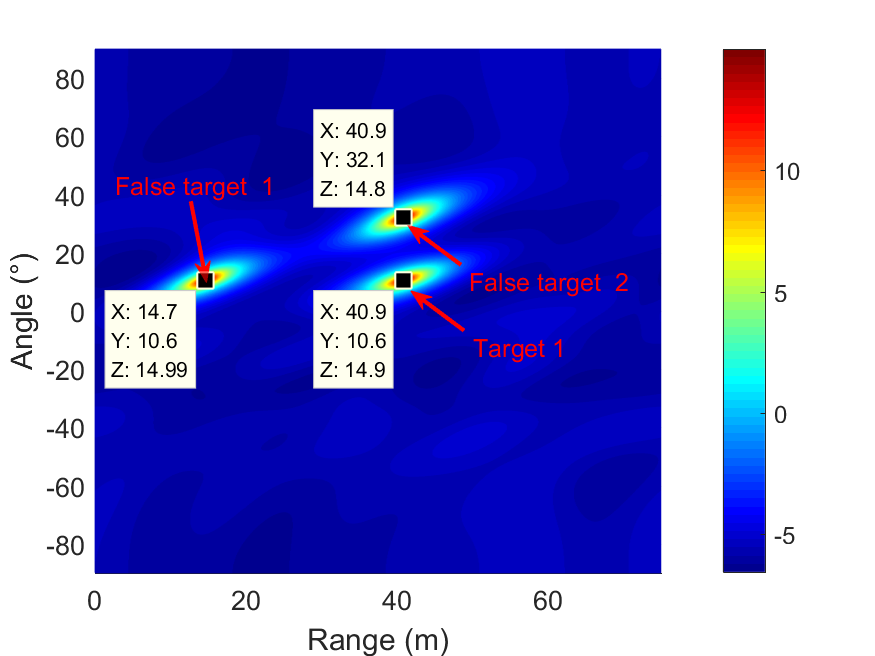}}\label{Spec_linear_deltaf1}
	\subfigure[]	{\includegraphics[width=0.24\textwidth]{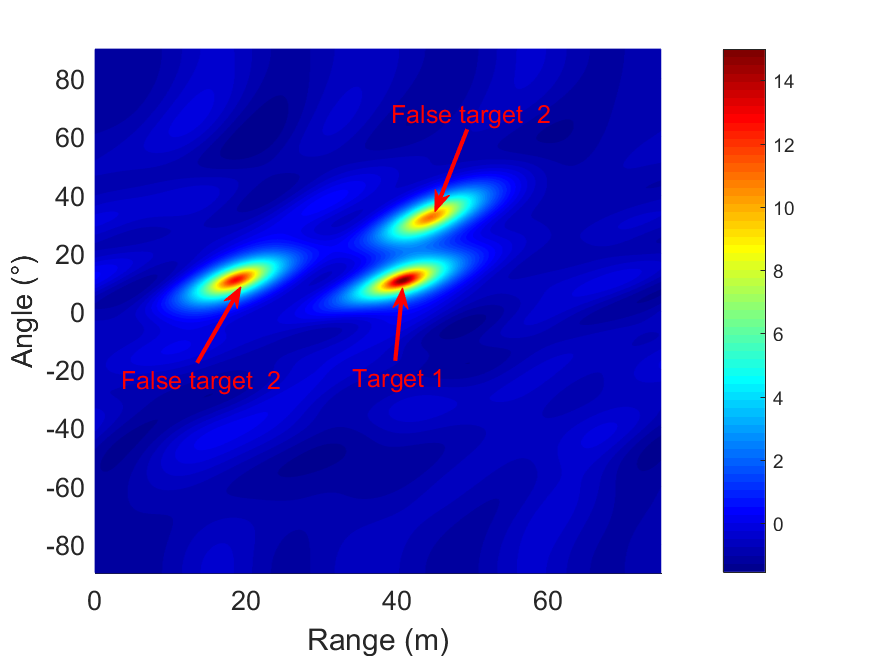}}\label{Spec_criteria_deltaf1}
	\caption{Spatial spectrum when estimate the target 1, where  $N=M=6$, $\mathrm{SNR}=5\mathrm{dB}$, $K=200$. (a) Signal transmit-receive space spectrum with LFO. (b) Signal transmit-receive space spectrum with FODC.}
	\label{Spatial_s}
	\vspace{-0.3cm} 
\end{figure}

To evaluate the proposed frequency offsets design criterion, in Fig. \ref{3D_amb1}, we compare the target recovery performance of the SSMTE and LCSSE algorithms under FODC and LFO. We set up 3 targets, as $\left\{ 10.55\degree,40.9\mathrm{m},8.62\mathrm{m}/\mathrm{s} \right\}$, $\left\{ 10.55\degree,89.6\mathrm{m},20.42\mathrm{m}/\mathrm{s} \right\}$, $\left\{ 32.01\degree,115.9\mathrm{m},36.5\mathrm{m}/\mathrm{s} \right\}$, respectively.  From Fig. \ref{3D_amb1} (a) and Fig. \ref{3D_amb1} (c), with LFO, one can observe that both SSMTE and LCSSE methods suffer from range ambiguity, where the targets' distances are estimated to other range bins, leading to incorrect parameter estimation and pairing. Fig. \ref{3D_amb1} (b) and Fig. \ref{3D_amb1} (d) show that the targets can be correctly estimated when using FODC. The reason for this benefit is shown in Fig. \ref{Spatial_s}. 

Fig. \ref{Spatial_s} gives the signal transmit-receive space spectrum with LFO and FODC. When adopting the linear frequency offset scheme, calling back to \eqref{a_T_R_g}, the range period of the transmit range steering vector is $\frac{c}{2\Delta f}=75$m. Therefore, when estimating a certain target, other targets will form false targets with high peaks in the spatial spectrum, as shown in \ref{Spatial_s}(a). This leads to target range estimation errors and target parameters pairing errors. In contrast, the proposed frequency offsets design criterion greatly increases the range period. Hence, targets in other range bins can hardly form peak values in the estimated range bin, reducing the probability of misestimation.

\begin{figure}[htbp]
	\centering
	\subfigure 	
	{\includegraphics[width=0.37\textwidth]{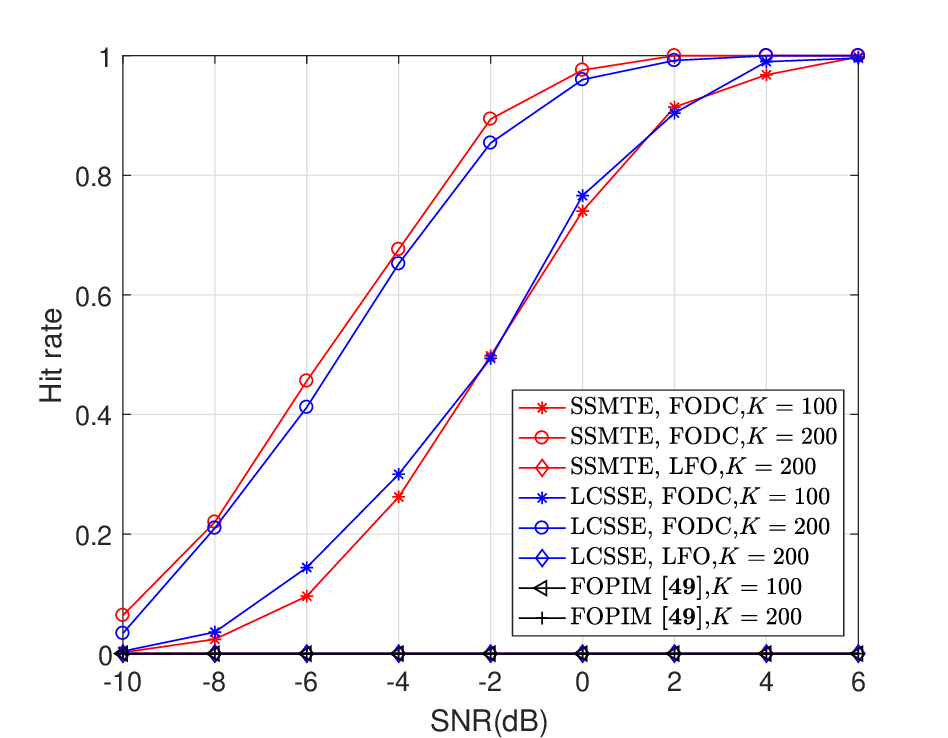}}
	\caption{Hit rate versus SNR with different snapshots, where $N=M=6$. }
	\label{Hit_rate_VS_SNR_Snaps}
\end{figure}

Fig. \ref{Hit_rate_VS_SNR_Snaps} compares the hit rates of the proposed methods with FOPIM and MIMO schemes. Note that the frequency offset pool size for the FOPIM scheme is set to $N$ \cite{jian2023fda}. Fig. \ref{Hit_rate_VS_SNR_Snaps} indicates that the hit rates of the proposed sensing approaches with FODC are improved by increasing the snapshots number. 
This is because more snapshots yield more accurate covariance matrix estimation results, thus improving the parameter recovery performance. 
Under the FODC method, the hit rate of the LCSSE method is approximately equal to that of the SSMTE method.
The hit rates remain 0 for SSMTE with LFO, LCSSE with LFO and FOPIM methods, indicating their inability to estimate parameters for multiple targets simultaneously. The phenomenon arises due to the distances of target 1 and target 3 differ by one period, causing range ambiguity and hindering parameter pairing in the FOPIM and LFO-ralated schemes.

\begin{figure}[htbp]
	\centering
	\subfigure[]	{\includegraphics[width=0.35\textwidth]{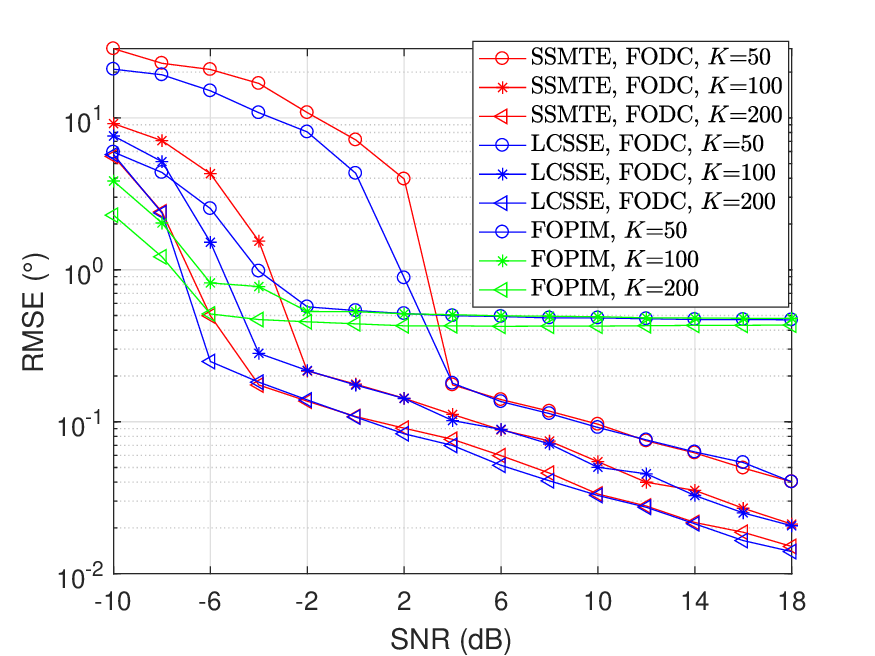}}\label{RMSE_comp_Ang_snaps}
	\subfigure[]	
	{\includegraphics[width=0.35\textwidth]{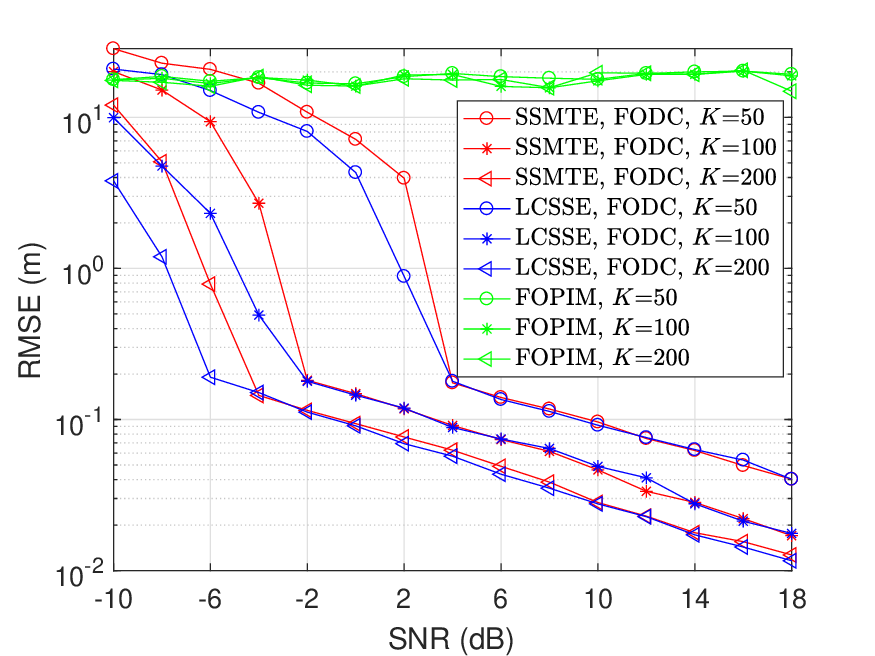}}\label{RMSE_comp_R_snaps}
	\caption{RMSE versus SNR with different number of snapshoots, where $N=M=6$. (a) Angle estimation. (b) Range estimation.}
	\label{RMSE_comp_snaps}
	\vspace{-0.3cm} 
\end{figure}

In Fig. \ref{RMSE_comp_snaps}, we compare the multi-target RMSE performance among different schemes. We find that the SSMTE and LCSSE methods have the similar range and angle estimation accuracies at middle to high SNRs. At very low SNRs, the LCSSE method is slightly better than the MSSTE method. This is because the angle and range errors of the SSMTE method are coupled. That is, in the low SNR region, the angle and range estimation errors are significant and affect each other. In contrast, in the LCSSE method, the targets range estimation does not affect the angle estimation. 
On the other hand, Fig. \ref{RMSE_comp_snaps}(a) shows that the angle estimation error of FOPIM stays around 0.45° after a brief drop, which is much higher than the proposed methods. This is attributed to the fact that the FOPIM method relies only on a simple receiver beamformer to estimate the angle, which has a very low angular resolution. There are two reasons why the FOPIM's range estimation error in \ref{RMSE_comp_snaps}(b) remains high: 1) the large angle estimation error causes a large distance estimation accuracy; 2) the FOPIM method is unable to pair multi-target angles and distances, resulting in mis-paired distance estimates.

\begin{figure}[htbp]
	\centering
	\subfigure[]	{\includegraphics[width=0.35\textwidth]{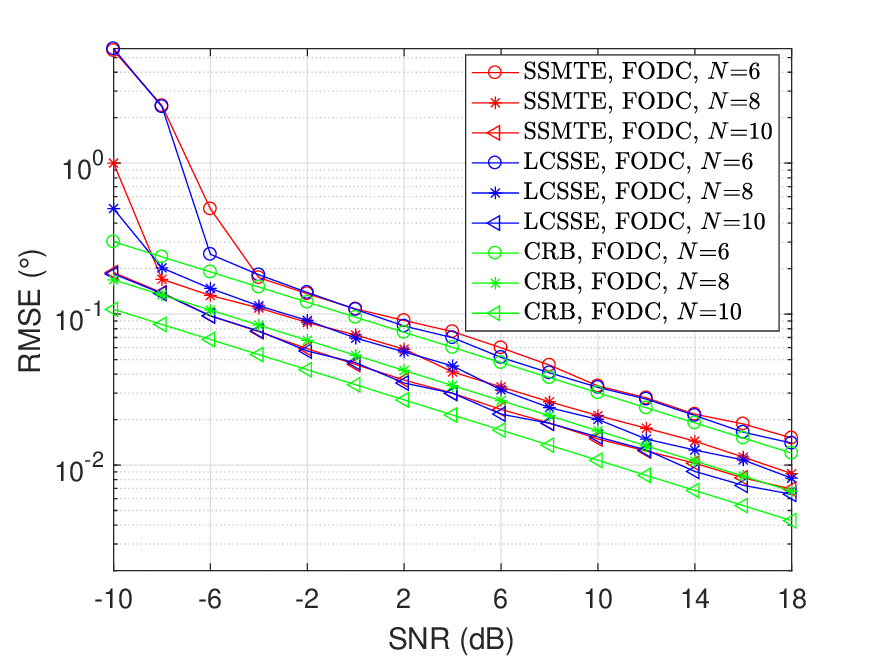}}\label{RMSE_ang_emit_antennas}
	\subfigure[]	
	{\includegraphics[width=0.35\textwidth]{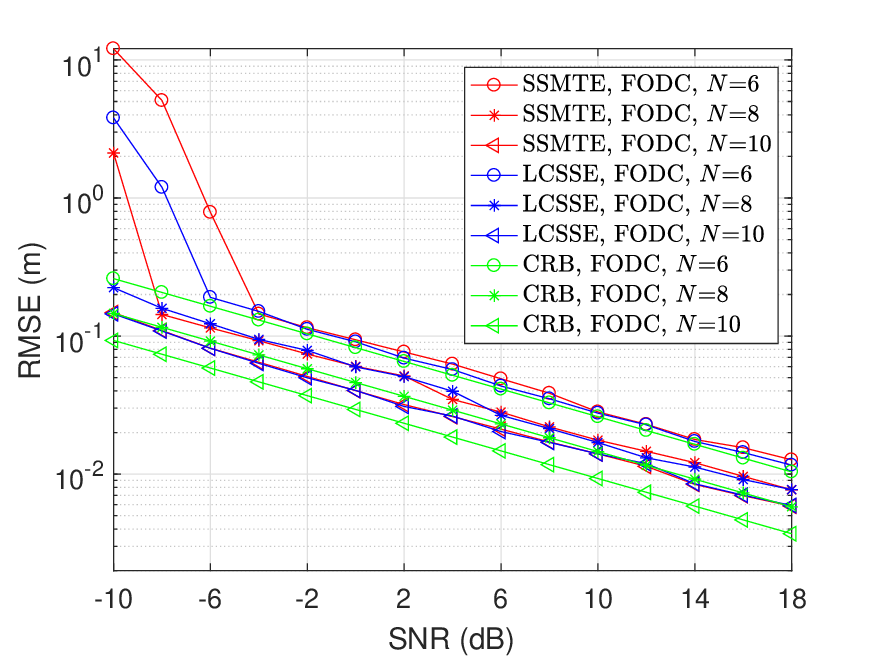}}\label{RMSE_R_emit_antennas}
	\subfigure[]	{\includegraphics[width=0.35\textwidth]{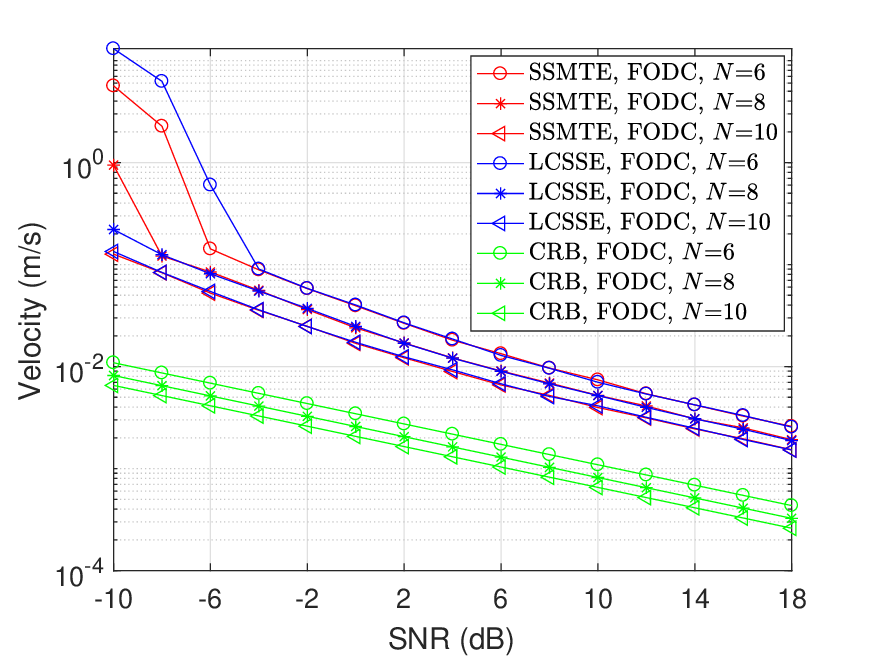}}\label{RMSE_vel_emit_antennas}
	\caption{RMSE versus SNR with different number of transmit and receive antennas, where $K=200$. (a) Angle estimation. (b) Range estimation. (c) Velocity estimation.}
	\label{RMSE_emit_antennas}
	\vspace{-0.3cm} 
\end{figure}
Fig. \ref{RMSE_emit_antennas} illustrates RMSEs and root CRBs for the proposed system with different numbers of transmit antennas when FODC is employed. Note that the frequency offsets are set as $\left\{\left\{ 0,1,2,3.17,4.2,5.2, 6.2, 7.2 \right\}\times \Delta f \right\} \mathrm{MHz}$ and $\left\{\left\{ 0,1,2,3.17,4.2,5.2, 6.2, 7.2, 8.2, 9.2\right\}\times \Delta f \right\} \mathrm{MHz}$ when $N=8$ and $N=10$, respectively. One can see that the estimation accuracy of target angle, range and velocity improves with the increasing number of transmit antennas. The SSMTE and LCSSE have the similar accuracies with different $N$. 
One can see that the angle and distance estimation performance is close to that of the CRB, but the velocity estimation differs significantly from the CRB. This is because in the proposed methods, the angle and distance estimation errors are substituted into the velocity estimation, which reduces the velocity estimation accuracy, while the velocity CRB is independent of the angle-distance estimation error. Nevertheless, at $N=6$, SNR=0dB, the velocity RMSE of the proposed method is 0.026m/s, which meets most civil sensing scenarios.

\begin{figure}[htbp]
	\centering
	\subfigure[]	{\includegraphics[width=0.37\textwidth]{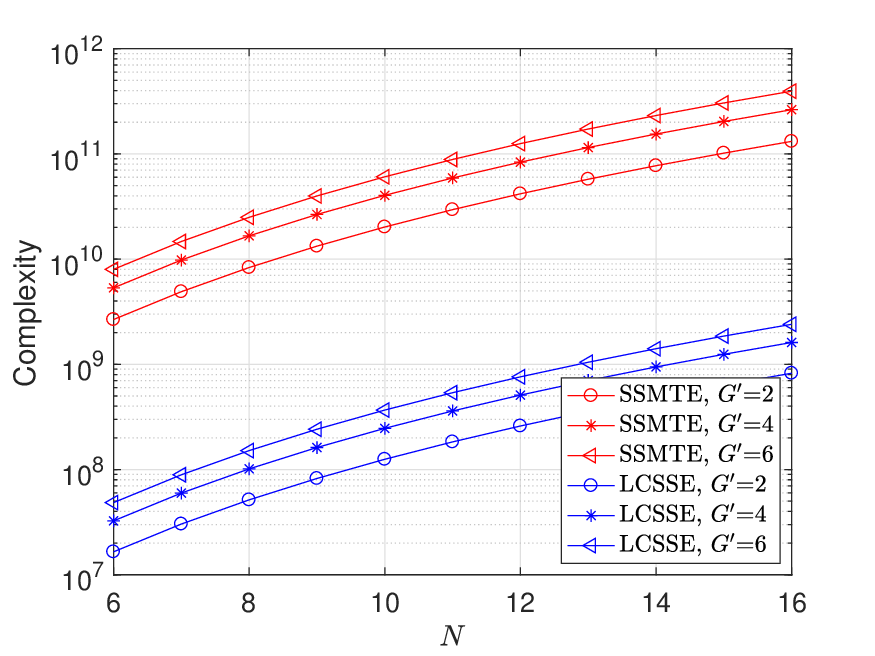}}
	\caption{Computation complexity versus the number of antennas, where $M=N$, $K=200$, $G=6$, $s_r=s_{\theta}=1000$.}
	\label{complexity_comp}
\vspace{-0.3cm} 
\end{figure}
Fig. \ref{complexity_comp} compares the computational complexity of the proposed SSMTE and LCSSE algorithms for different $N$. The complexity of both methods increases with the number of range bins ($G'$) to be estimated. The complexity of the LCSSE is two orders of magnitude lower than that of the SSMTE method, thanks to its conversion of a 2-D angle-range joint search into two 1-D searches. Considering the sensing performance comparisons in Fig. \ref{Hit_rate_VS_SNR_Snaps} to Fig. \ref{RMSE_emit_antennas}, we conclude that the LCSSE approach is the wiser choice to sense targets in the proposed system.

\subsection{Communication Simulation}
In this subsection, we investigate the communication performance of the proposed CCIE scheme. Note that ``Ana" and ``Sim" represent the BER theoretical upper bound and the Monte Carlo simulation results in the following figures, respectively.

\begin{figure}[htbp]
	\centering
	\subfigure[]	
	{\includegraphics[width=0.32\textwidth]{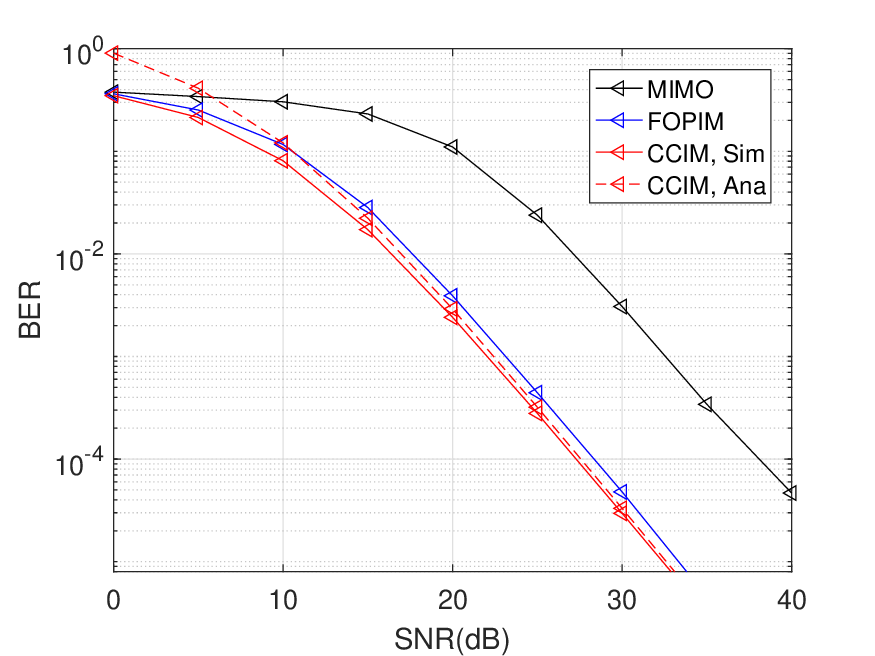}}\label{BER_Rx_4N_2U}
	\subfigure[]	{\includegraphics[width=0.32\textwidth]{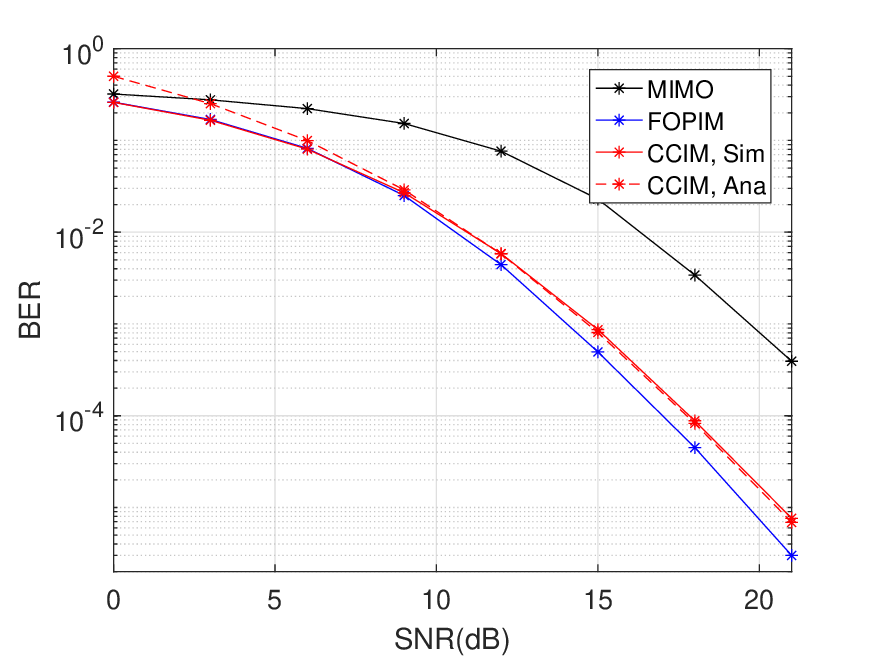}}\label{BER_Rx_4N_4U}
	\caption{BER versus SNR with different number of receive antennas, where $N=4$, $J=4$, $L=4$. (a) $U=2$. (b) $U=4$. }
	\label{BER_diff_Rx_antenas}
	\vspace{-0.3cm} 
\end{figure}
Fig. \ref{BER_diff_Rx_antenas} compares the BER of the proposed CCIE method with FOPIM \cite{jian2023fda} and traditional MIMO \cite{biglieri2007mimo} methods for varying number of receive antennas. Note that the CCIE method carries 16 bits, and for fairness, the frequency offset pool size for FOPIM is set to 4, and the modulation order for both the FOPIM and MIMO methods is set to 8. From Fig. \ref{BER_diff_Rx_antenas}, we observe that the system BER is improved with the increasing $U$, which stems from the fact that the higher receive diversity gain. The simulations of the CCIE method match well with the theoretical results, which verifies the BER analysis. Moreover, MIMO shows the worst BER performance among the three schemes. This is because its transmit symbols are coupled to each other, resulting in a small judgment domain, which deteriorates the BER performance. 

Another interesting finding is that the CCIE method outperforms the FOPIM method when the number of receiving antennas is small ($U$=2). However, when $U$ increases, the BER of the proposed CCIE method is worse than FOPIM. This can be explained as follows, referring Eq. (38) in \cite{jian2023fda} gives that the frequency offset permutation estimation error probability ($P_{\mathrm{perm}}$) of the FOPIM scheme is governed by the frequency offset combination estimation error probability ($P_{\mathrm{comb}}$). $P_{\mathrm{comb}}$ remains high with low $U$, resulting in a high overall index bit error probability. As $U$ increases, $P_{\mathrm{comb}}$ decreases dramatically. Furthermore, comparing (42) in \cite{jian2023fda} and \eqref{kappa_var} in this paper reveals that the judgment domain spacing of the FOPIM method is larger than that of CCIE, which results in a lower BER for FOPIM than for CCIE when larger $U$.

\begin{figure*}[htbp]
	\centering
	\subfigure[]	
	{\includegraphics[width=0.32\textwidth]{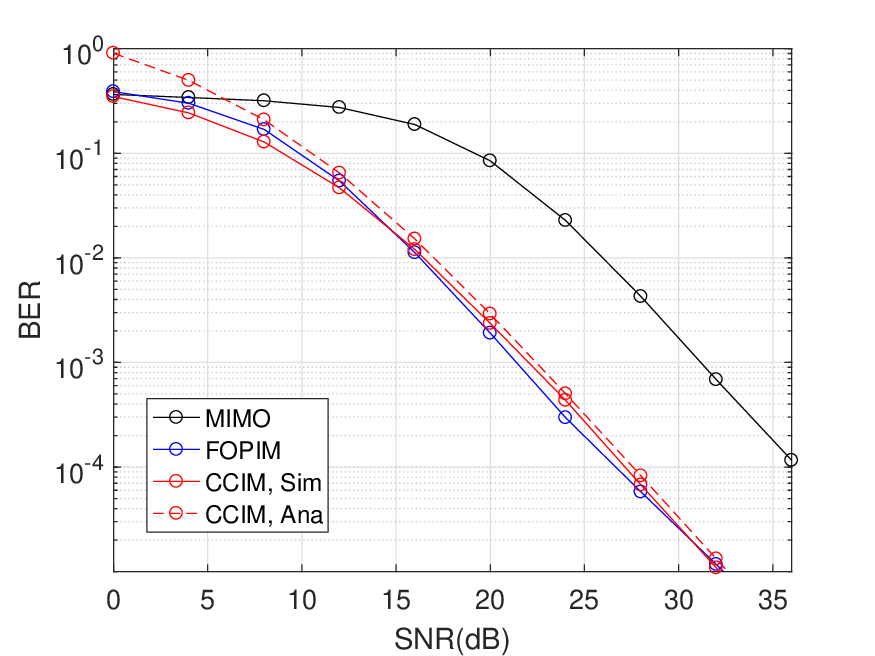}}\label{BER_Rx_6N_2U}
	\subfigure[]	{\includegraphics[width=0.32\textwidth]{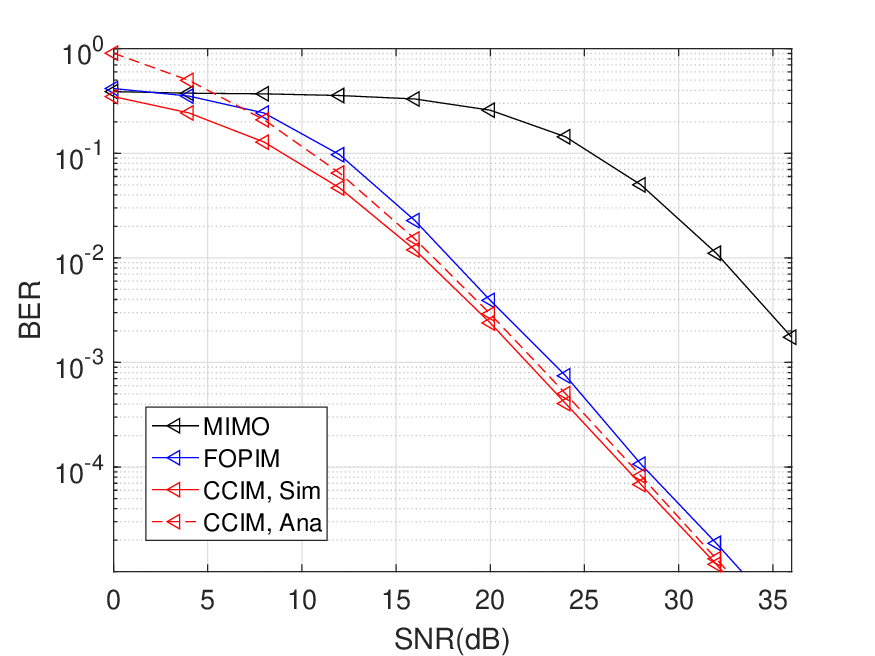}}\label{BER_Rx_8N_2U}
	\subfigure[]	{\includegraphics[width=0.32\textwidth]{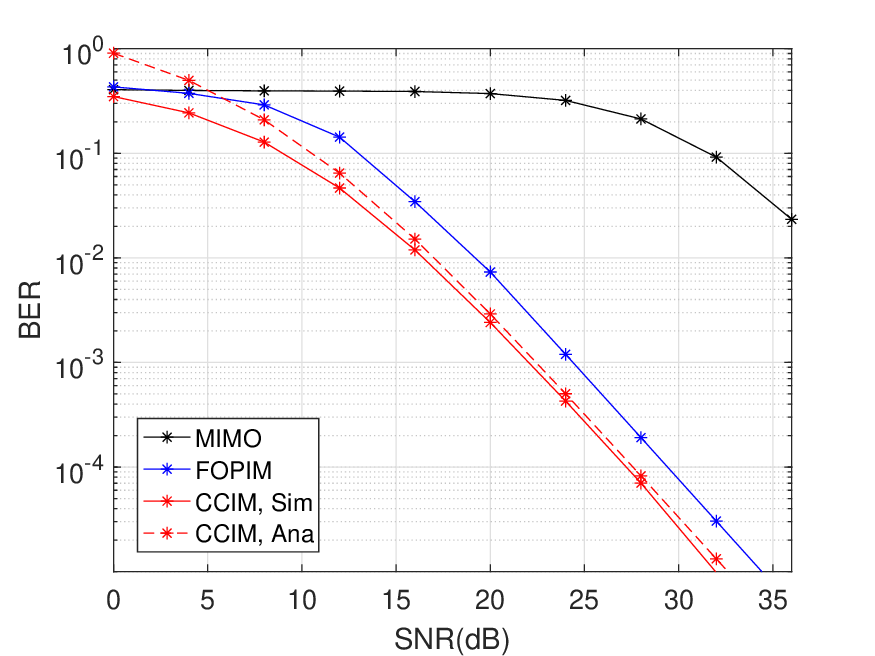}}\label{BER_Rx_10N_2U}
	\caption{BER versus SNR with different number of transmit antennas, where $J=4$, $U=2$, $L=4$. (a) $N=6$. (b) $N=8$. (c) $N=10$.}
	\label{BER_Rx_Ns_2U}
	\vspace{-0.3cm} 
\end{figure*}

Fig. \ref{BER_Rx_Ns_2U} shows the BER comparison results with different number of transmit antennas. Note that the parameter configurations for MIMO and FOPIM methods are the same as for the CCIE scheme. As $N$ increases, the BER performance of the CCIE scheme gradually outperforms that of the FOPIM scheme. Moreover, the BER of the FOPIM and MIMO approaches increase with increasing $N$, whereas the CCIE's BER  remains with increasing $N$. This is because as $N$ increases, the FOPIM method suffers a higher error probability in estimating frequency offsets. On the other hand, \eqref{y_n_k_bf} gives that bits carried by every transmit antenna are decoded independently in the CCIE method, with no dependence on $N$. Therefore, we conclude that CCIE can achieve higher communication rates without loss of BER performance by increasing the number of transmit antennas.

\begin{figure}[htbp]
	\centering
	\subfigure 	
	{\includegraphics[width=0.37\textwidth]{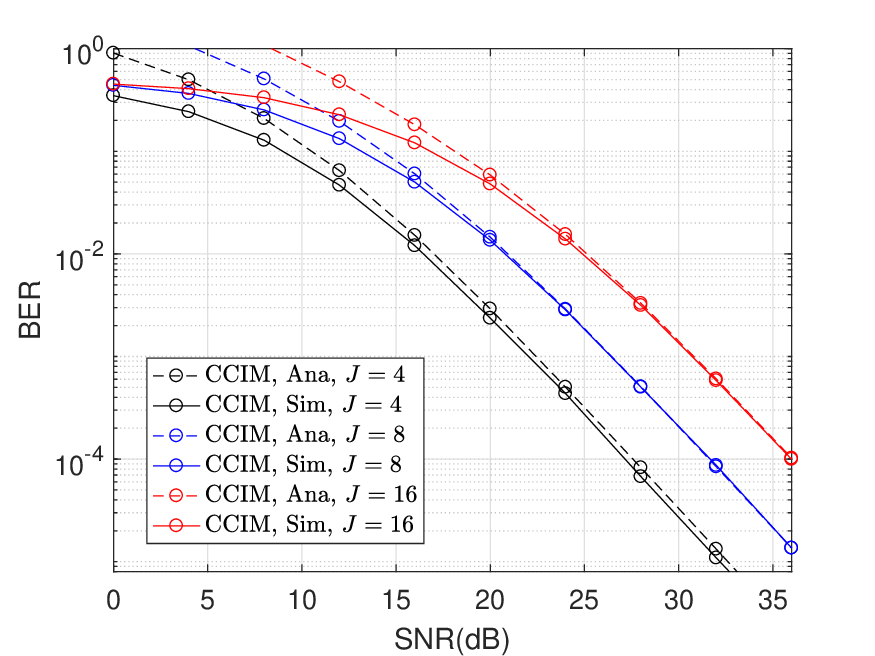}}
	\caption{BER versus SNR with varying size of complex coefficient vector, where $N=6$, $U=2$, $L=4$.}
	\label{BER_setsize}
	\vspace{-0.3cm} 
\end{figure}
In Fig. \ref{BER_setsize}, we study the BER of the CCIE scheme with the different size of the complex coefficient set. It is seen that the BER of the proposed CCIE approach rises as the size of the complex coefficient set becomes larger. For every doubling of $J$, the BER performance decreases about 4 dB. The reason for this phenomenon can be found in \eqref{P_IM}, where the index bits misestimation probability $P_{IM}$ increases with increasing $J$, leading to a deterioration in the system BER performance.

\begin{figure}[htbp]
	\centering
	\subfigure 	
	{\includegraphics[width=0.37\textwidth]{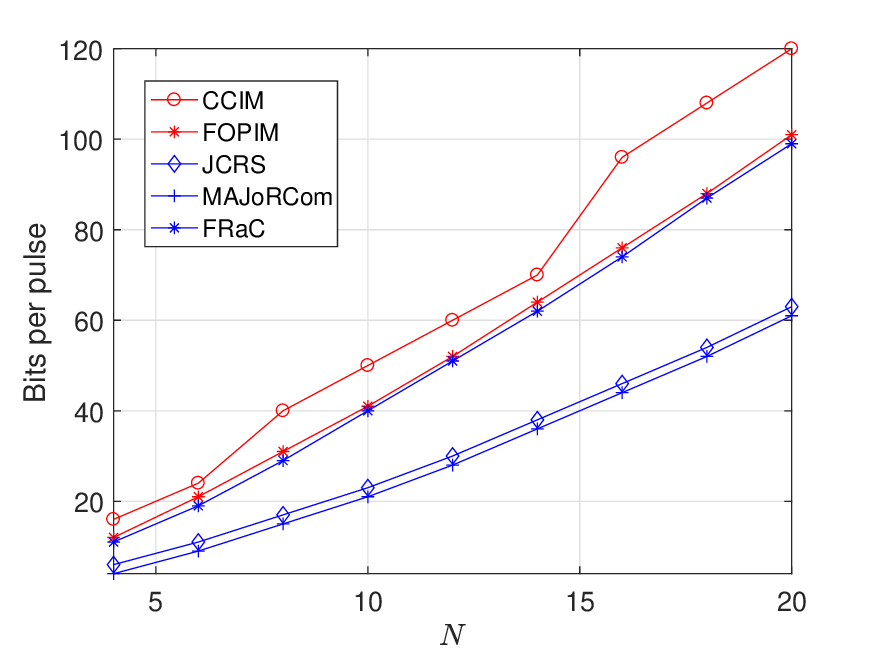}}
	\caption{Bits per pulse comparison among different ISAC schemes with varing number of transmit antennas, where $L=4$.}
	\label{rates_comp_diff_N}
	\vspace{-0.3cm} 
\end{figure}
Fig. \ref{rates_comp_diff_N} compares the bits per pulse among different ISAC schemes: proposed CCIE, FOPIM \cite{jian2023fda}, FRaC \cite{ma2021frac}, JCRS \cite{li2023joint}, MAJoRCom \cite{huang2020majorcom}. In the simulation, the total bandwidth of FOPIM is set equal to that of CCIE, namely, the size of FOPIM's frequency offset pool is set to $N$. To be fair, we set $J=N$ for CCIE to have the same index resource. JCRS has a waveform set size equal to $J$. MAJoRCom uses a separate frequency for each antenna, while FRaC activates $N-2$ antennas. Studying Fig. \ref{rates_comp_diff_N} finds that FOPIM scheme carries more bits than FRaC, JCRS and MAJoRCom schemes, the reason for this phenomenon has discussed in \cite{jian2023fda}. Moreover, Fig. \ref{rates_comp_diff_N} depicts that the proposed CCIE method outperforms the FOPIM approach in terms of bits per pulse performance. This observation can be elaborated as follows: in Fig. \ref{rates_comp_diff_N} the bits per pulse for CCIE and FOPIM are $N\times (\lfloor \log _2N \rfloor +\log _2L)$ and $N\log _2L+\lfloor \log _2N! \rfloor +\lfloor \log _2C_{N}^{N} \rfloor$, respectively. Since $N^N>N!$, we see that CCIE carries more bits than that FOPIM method.

\section{Conclusion}
\label{sc6}
This paper investigated the FDA-MIMMO-based ISAC system in a multi-target sensing scenario. \textcolor{black}{Specifically, the SSMTE method was proposed to estimate targets. The angles and distances of targets were estimated by 2-D search of the target-containing spatial spectrum.} Then, the targets velocities were estimated by the LS method. To reduce the complexity, we designed the LCSSE method to reduce the complexity by converting the 2-D search into two 1-D searches. On the other hand, the FDA-MIMO's range steering vector suffered range ambiguity. To address this issue, the FOCD scheme was proposed, which adjusted the integer and fractional parts of each transmit frequency offset to enlarge the range periodicity, thereby mitigating range ambiguity in multi-target estimation. \textcolor{black}{ Moreover, to improve the communication rate, a CCIE scheme was proposed at the transmitter, which carried the extra bits by selecting complex coefficients. Besides, the closed-form expressions for CRB, complexity and BER upper bound are derived. Simulation results illustrated that the LCSSE method dramatically reduced the complexity of SSMTE with no degradation in sensing accuracy. Moreover, the proposed FDA-MIMO-based ISAC system outperforms the FOPIM based ISAC system in terms of multi-target sensing performance. }

\appendices

\ifCLASSOPTIONcaptionsoff
\newpage
\fi

\bibliographystyle{IEEEtran}
\bibliography{ref}

\end{document}